\documentclass[aps,prstab,onecolumn,superscriptaddress,showpacs]{revtex4}
\usepackage{graphicx}
\usepackage{url}
\usepackage[colorlinks,linkcolor=blue,anchorcolor=blue,citecolor=blue]{hyperref} % hyper reference to contents 
\usepackage{multirow}
\usepackage{color}

\newcommand {\htp}{$\text{H}_2^+$~}

\include{mylatexdefs}

\begin{document}

\title{Beam Dynamics Simulation for the High Intensity DAE$\delta$ALUS Cyclotrons}

\author{J. J. Yang}
\email{yangjianjun2000@tsinghua.org.cn}
\affiliation{Massachusetts Institute of Technology, Cambridge, MA 02139, USA}
\affiliation{Paul Scherrer Institut, Villigen, CH-5234, Switzerland}
\affiliation{China Institute of Atomic Energy, Beijing, 102413, China}
\author{A. Adelmann}
\email{andreas.adelmann@psi.ch}
\affiliation{Paul Scherrer Institut, Villigen, CH-5234, Switzerland}
\author{W. Barletta}
\affiliation{Massachusetts Institute of Technology, Cambridge, MA 02139, USA}
\author{L. Calabretta}
\affiliation{Istituto Nazionale di Fisica Nucleare, Laboratori Nazionali del Sud, I-95123, Italy}
\author{J. M. Conrad}
\affiliation{Massachusetts Institute of Technology, Cambridge, MA 02139, USA}

\noaffiliation
\begin{abstract}
 In the DAE$\delta$ALUS (Decay-At-rest Experiment for $\delta_{CP}$ studies At the Laboratory for Underground Science) project, 
 high power \htp cyclotron chains are proposed to efficiently provide proton beams with a kinetic energy of 800 MeV and an average power in the MW range.  
 Space charge plays a pivotal role in both the injector and the ring cyclotrons. 
Large-scale particle simulations show that the injector cyclotron is a space charge dominated cyclotron and that a 5mA beam current can be  extracted with  tolerable beam losses on the septum.
In contrast, in the ring cyclotron, no space charge induced beam loss is observed during acceleration and extraction. 
\end{abstract}


\maketitle

\section{Introduction \label{intro}}
The DAE$\delta$ALUS collaboration is designing advanced cyclotrons that accelerate molecular hydrogen ions to produce
decay-at-rest neutrino beams for a novel search for $CP$ violation in the neutrino sector{~\cite{Conrad:1}}.  
The DAE$\delta$ALUS experiment calls for measurements at three values: L = 1.5 km, 8 km and 20 km. 
Three accelerator modules, called near-, mid- and far-site are used, providing 800 MeV protons with 0.8, 1.6 and 4.8 MW average beam power respectively.  
The experiment imposes a strong constraint on each accelerator in the complex, that is, to operate with a 20\% duty cycle
that allows the identification of the source of a neutrino event, and the measurement of the cosmogenic backgrounds \cite{Alonso:2010fs}.
Consequently, the current during beam-on should be 5 times the average beam current; i.e., 
in the mid-site accelerator module, 2 mA of average current requires 10 mA of peak current.
The additional constraints on the accelerators of compactness and cost-effectiveness are not addressed in this paper.

\begin{figure}[ht!]
\begin{center}
{\includegraphics[width=0.8\linewidth]{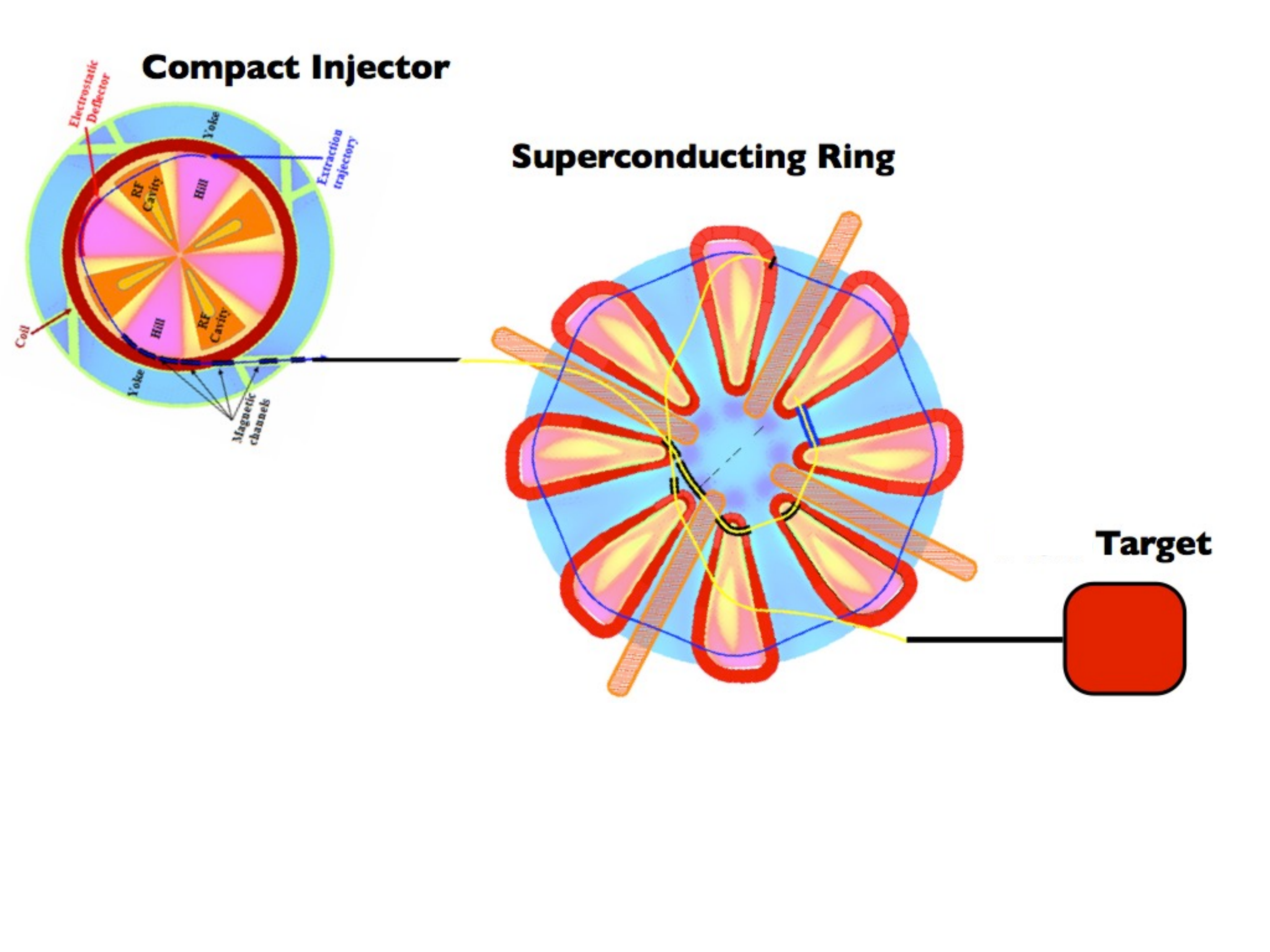}}
\end{center}
\caption{A schematic overview of the accelerator complex at the mid-site accelerator module, 
which is located at 8 km away from the underground detector. Only one extraction channel from the DSRC is shown in this schematic drawing.
\label{layout}}
\end{figure}

This paper describes the beam dynamics study of the mid-site cyclotrons.
The machine configuration discussed here, shown schematically in Fig.~\ref{layout}, is the mid-site cyclotron complex consisting of two cascaded cyclotrons. 
The injector cyclotron (DIC) is a four-sector compact machine, which accelerates a beam of \htp up to 60 MeV/amu. 
The beam is then extracted by an electrostatic deflector and is transported and injected into an eight-sector superconducting ring cyclotron (DSRC), 
in which  the beam is accelerated to 800 MeV/amu by four single-gap rf-cavities.
Two stripper foils can be used to extract two proton beams at the same time from the ring cyclotron, however, in Fig.\ 1 for sake of simplicity just one trajectory is shown.
The key parameters of these two cyclotrons are listed in Table \ref{tab:cycs}.
The general characteristics of all the sub-systems are elaborated in the Ref.{~\cite{EricePaper,Luciano:1, Luciano:2}}.

\begin{table*}
\caption{\label{tab:cycs} Key parameters of the  mid-site cyclotrons}
\begin{tabular}{lcccccccccc}
\hline \hline
& type & orbit radius & kin. energy &  avg. power & avg. field & harmonic&sector& cavity & cavity &  turn\\
& &(cm) & (MeV/amu) & (MW) & (T) & no. &no. & type & no.& no.\\
\hline
DIC & normal &5\dots200 & 0.035\dots60 & 0.12 & 0.95\dots1.17 & 6 & 4 & double-gap&4  &107 \\
DSRC & superconducting& 190\dots480 & 60\dots800&1.6 & 1.06\dots1.88 &6 & 8 & single-gap& 4 & 401 \\
\hline \hline
\end{tabular}
\end{table*}

One of the main challenges of this high power accelerator complex concerns the control of space charge effects and the beam loss rates at extraction for both cyclotrons. 
The choice of accelerating \htp  instead of protons has two major advantages:
a) The extraction of \htp in the DSRC with a stripping foil does not require a clean turn separation at extraction, which would be mandatory for bare protons.
b) In addition the space charge effects for \htp are smaller. A measure of the strength of space charge is the generalized perveance~\cite{reiser2008theory}
\begin{equation} \label{eq:per}
K = \frac{qI}{2\pi\varepsilon_0 m \gamma^3 \beta^3},
\end{equation}
where $q, I, m, \gamma$ and $\beta$ are respectively the charge, current, rest mass and relativistic
parameters of the particle beam. 
The higher the value of $K$, the stronger the space charge effects are.
According to Eq.(\ref{eq:per}) the $K$ factor for 5 mA, 800 MeV/amu \htp beam in the DSRC is equal to that of 2.5 mA, 
800 MeV proton beam with the same  $\gamma$. 
In consequence it is similar to the space charge effects that are present in the 2.4 mA, 590 MeV proton ring cyclotron at PSI.
Similarly,  the perveance of a 5 mA, 35 keV/amu \htp in the DIC is approximately equal to that of a 2 mA proton beam at 30 keV, 
which has already been achieved in commercial compact cyclotrons, such as CYCLONE-30 \cite{IBA} and TR-30 \cite{cyclotron-solutions}. 
Therefore, a 5 mA \htp beam seems not far beyond contemporary cyclotron capabilities with respect to space charge.

This simple analysis, however,  offers only the simplest understanding about space charge effects.
Many other factors, such as the external focusing, the path length in cyclotrons, 
have significant impacts on the overall effects of space charge and the associated beam dynamics. 
Therefore, one can not get a complete picture without precise calculations of beam dynamics.   
A practical project demands detailed prediction of the effects of space charge on the beam's evolution
based on accurate beam dynamics simulations of these high power cyclotrons.
The stringent systems requirements of limiting controlled and uncontrolled particles losses requires simulations and measurements with more than three to four orders of magnitude in dynamic range. 
Simulation and modeling using large scale optimization techniques
will be future research directions to meet the stringent requirements with respect to precise models and online control.

%For high intensity accelerator, it is crucial to study the the space charge and related beam loss from the very beginning of a project. 
The space charge effects in the DAE$\delta$ALUS cyclotrons have been studied quantitatively by numerical methods implemented in the parallel code  OPAL~\cite{Adelmann:1}.   
The beam dynamics model used in the simulation is described in Ref.~\cite{Yang:1,Yang:2}. 
The report of the first design \cite{Luciano:2} has been followed by several iterations of magnet design and space charge simulations.
Accordingly, the cyclotron structures have been improved to reduce beam loss to the acceptable range.
The following sections discuss the space charge simulations based on the latest design solution of both DIC and DSRC \cite{EricePaper}. 

\section{ Space charge in the DIC cyclotron\label{DIC}}
The maximum beam power delivered by the injector is determined mainly by space charge effects.  
Space charge forces can increase the beam emittance and consequently reduce the extraction efficiency. 
Reduced efficiency could be a serious limitation because beam losses must be kept below 200 W. 
This limit, following the experience of PSI and other laboratories, 
is found to be a practical maximum to allow for routine hands-on maintenance in the cyclotron vault. 
For the DIC, 200 W corresponds to an allowed beam loss of 1.67 $\mu$A of \htp beam at extraction.
Hence one must pay close attention when designing for clean extraction from the injector cyclotron.

\subsection {Stationary matched beam formation}
As is described in literature \cite{Gordon:1, kleeven:1, Adam:0, Bert:2001, Ada:1, Pozdeyev:2}, 
the  space charge force combined with the  radial-longitudinal coupling motion develops a vortex motion inside the bunch
that can change the shape of the beam.
%an intense particle beam,  properly matched to a cyclotron, develops a spatial stationary circular bunch distribution. 
%Consequently the halo is significantly reduced and a corresponding longitudinal decrease of the beam size is observed. The resulted
%compact beam with a minimal energy spread then an be extracted with minimal losses. 
If the  vortex motion  in the horizontal-longitudinal plane is strong enough, the beam will be approximately circular in this plane, 
as shown in \cite{Adam:2, Yang:1}  and experimentally verified \cite{HumbPC} at PSI.  
In the PSI Injector II cyclotron  a stationary compact beam is developed within the first several turns, 
which remains essentially unchanged until extraction and the beam phase width is about $2^\circ$ at the last turn.  
Therefore, the two flattop cavities of this cyclotron, that were originally designed to suppress the increase of energy spread for the long beam,  
are now operating as accelerating cavities. With this setting, Injector II can deliver up to 3 mA of proton beam for the PSI Ring Cyclotron. 
To evaluate the necessity of the flattop cavity for the DIC cyclotron, one must investigate whether this working mode exists.

When a coasting beam is perfectly matched with the cyclotron,  
its beam size oscillates periodically with a frequency equal to the number of sectors.
Baumgarten  \cite{Baumgarten:1} developed a theoretical model to compute the second moments of an ellipsoidal distribution ($\sigma$ matrix) matched to a cyclotron,
in a linear approximation for a given beam current and emittance.\
In practice, the beam does not necessarily maintain an uniform distribution. Hence there exists non-linear components that we have to take into account.\
The full 3D code OPAL provides us with numerical methods to compute not only  the $\sigma$ matrix, 
but also realistic particle distributions of a matched beam which includes nonlinear space charge forces.  

\begin{figure}[ht!]
   \centering
 {\includegraphics[width=0.4\linewidth,trim=2.0cm 2cm 2.0cm 2cm]{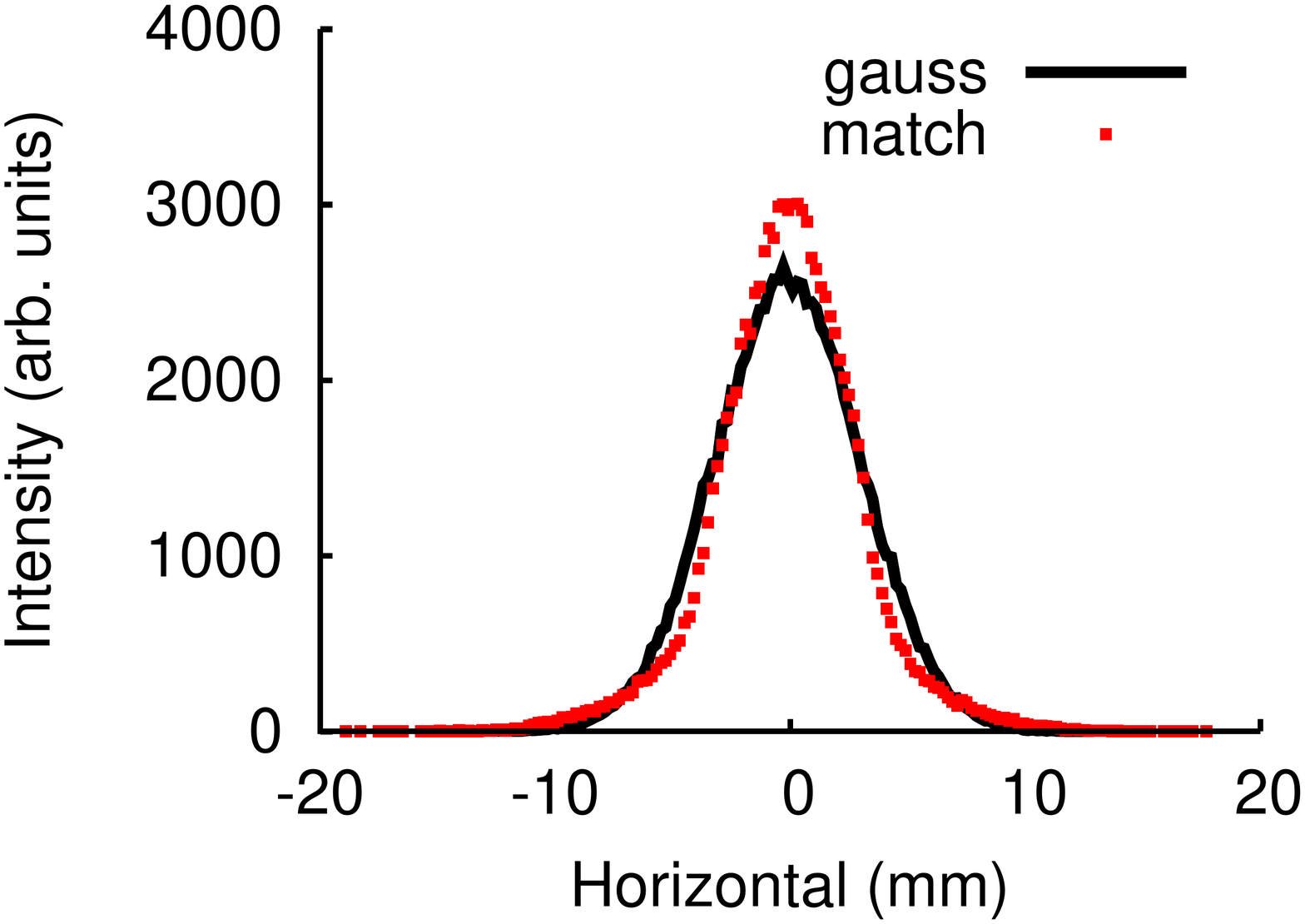}}
  {\includegraphics[width=0.4\linewidth,trim=2.0cm 2cm 2.0cm 2cm]{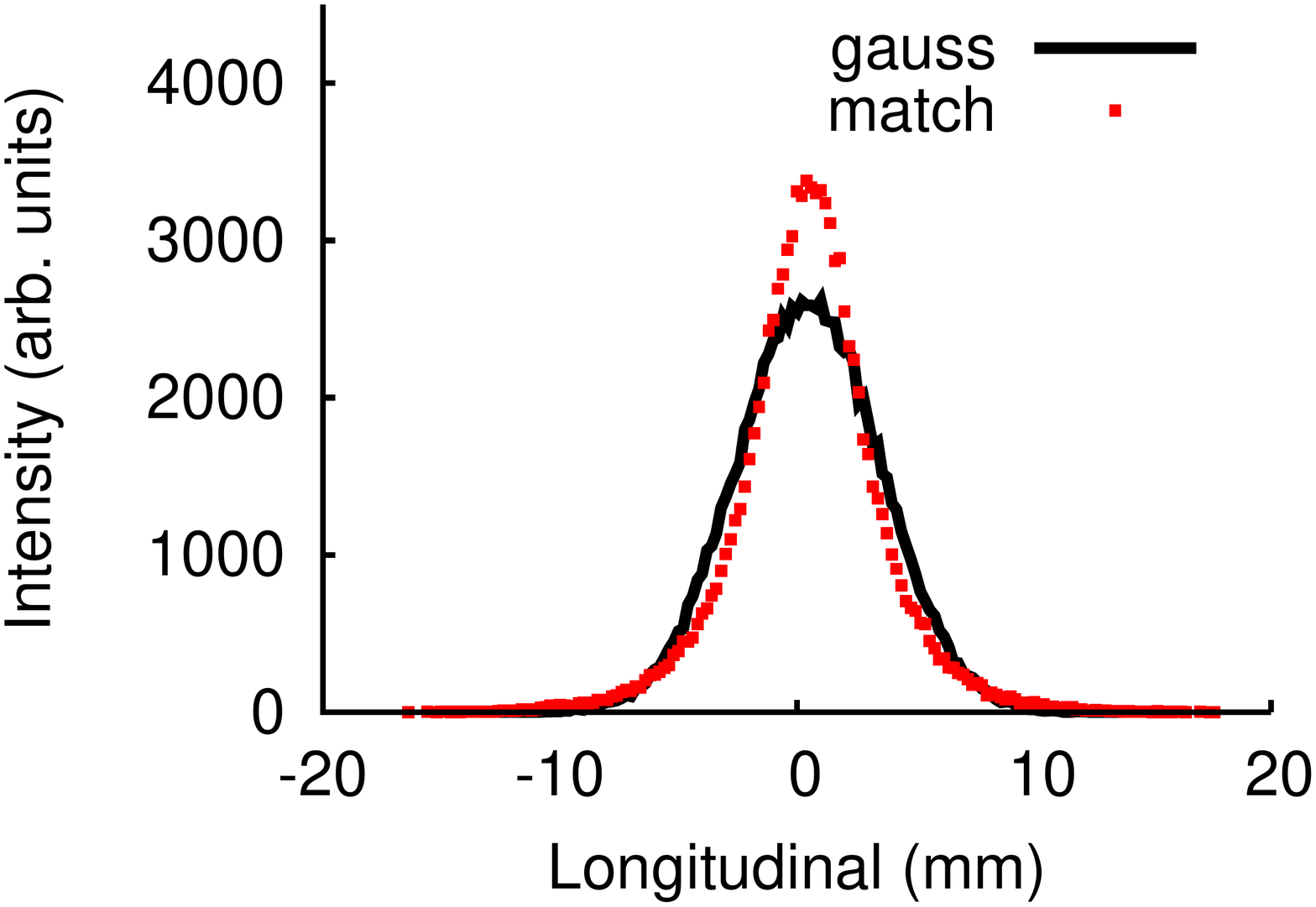}}
\caption{The longitudinal beam profile of the matched distribution and its Gaussion-fit with the same rms size }%for 10mA current and 1.5MeV/amu coasting beam}
    \label{fig:d7}
  \end{figure}

To search for a matched distribution, including non-linear space charge forces, we use as initial guess a spherical symmetrical particle
distribution as suggested in Ref.\ \cite{Baumgarten:1} with the beam radius expressed by 
\begin{equation} \label{eq:size}
\sigma=\left\{ \begin{array}{lll} 
\sigma_0 \left( 1+\frac{\alpha}{4}-\frac{\alpha^2}{32} \right) & \mbox{for} & 0\leq \alpha < 5/2 \\ 
\sigma_0 \left( 1+\alpha \right)^{1/3} & \mbox{for} & 5/2 \leq \alpha 
\end{array}\right. ,
\end{equation}
together with
\begin{equation} \label{eq:alpha}
\begin{array}{cc} 
\alpha = \frac{q\mu_0 I}{5\sqrt{10}mc\gamma h}\left(\frac{a}{\epsilon_n}\right)^{3/2} \text{ and } &\sigma_0 =\frac{\sqrt{2a\epsilon_n}}{\gamma},
\end{array}
\end{equation}
where $c$ is the speed of light, $\epsilon_n$ the normalized emittance, $a=c/\omega_0$ the cyclotron radius, $I$ the beam  current, $h$ the harmonic number, and $\mu_0$ vacuum permeability.
 \begin{figure}[ht!]
   \centering
 {\includegraphics[width=0.47\linewidth,trim=2.0cm 2cm 2.0cm 2cm]{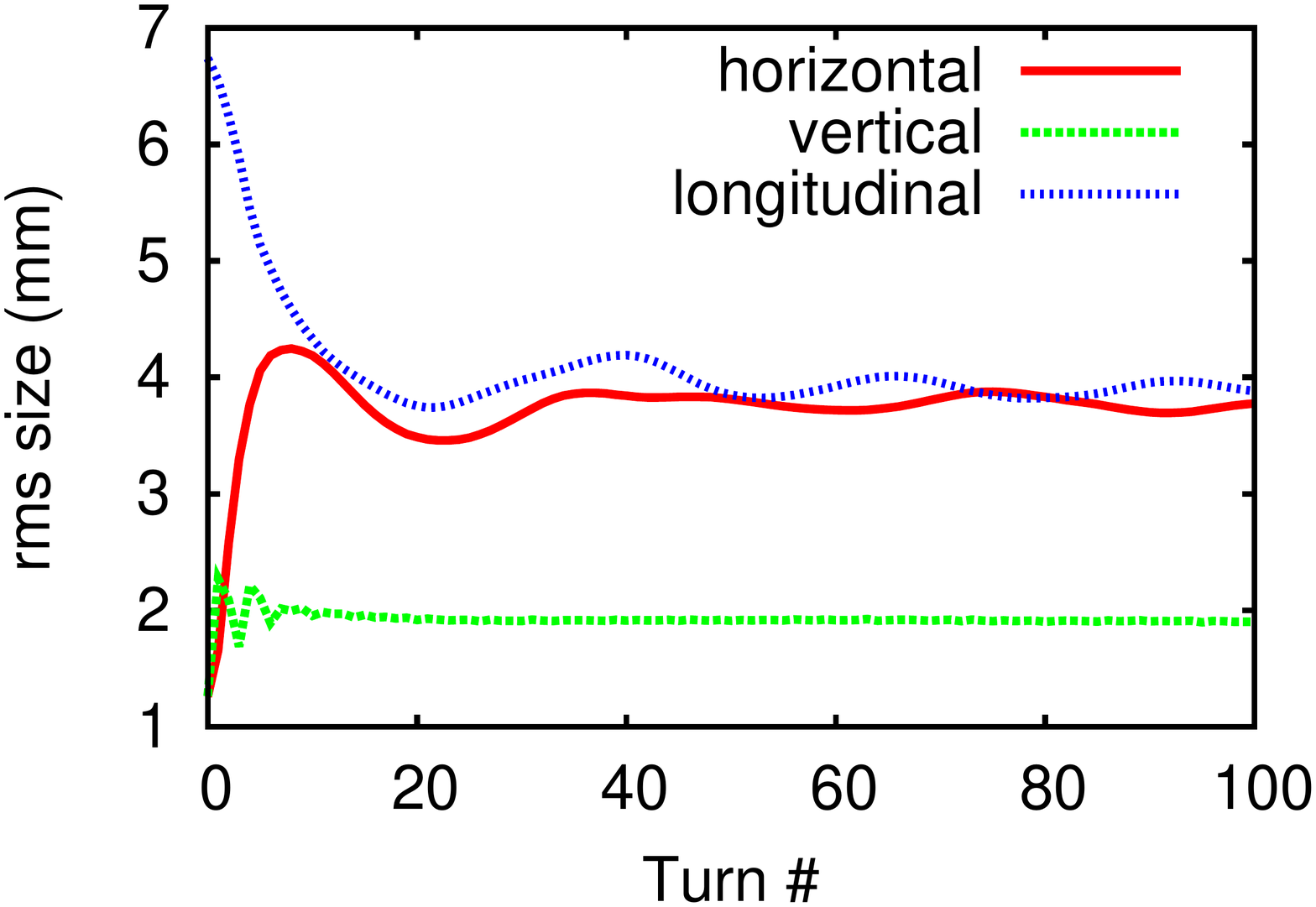}}
  {\includegraphics[width=0.47\linewidth,trim=2.0cm 2cm 2.0cm 2cm]{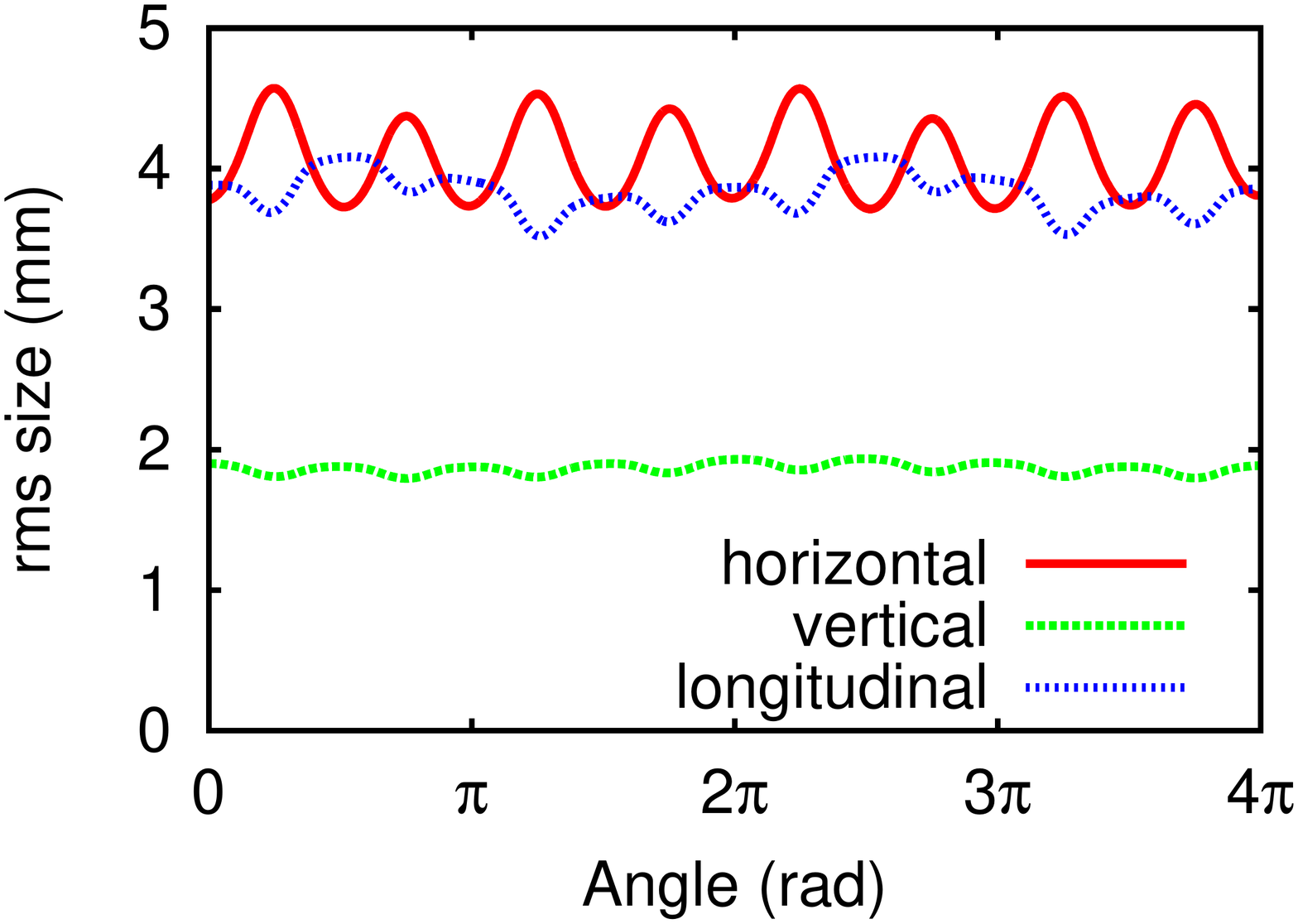}}
\caption{ The rms size snapshot at $0^\circ$ azimuth in the 100 turns (left), and the rms envelop in the last two turns (right). For this simulation, $10^5$  particles and mesh of $32^3$ grid points is used.}
    \label{fig:d6}
  \end{figure}
We start by tracking a coasting beam with 5 mA, 1.5 MeV/amu until the equipartitioning process has resulted in a matched, stationary distribution. 
The tracking result shows that the sizes of the initial unmatched beam converge to almost constant values within approximately 50 turns. 
The longitudinal beam size approaches the horizontal size, as is shown in Fig.\ \ref{fig:d6} (left). 
% yang: right figure is not a zoom at 0 degree, but the envelop for all the azimuth within 99th and 100 turns.
At different azimuths for the last two turns, the beam size oscillates with a frequency of four per turn (equal to the sector number of the cyclotron), 
as is shown in Fig.\ \ref{fig:d6} (right).   This behavior indicates that the stationary matched distribution has already developed. 
Figure \ref{fig:d7} shows the horizontal and longitudinal profile of the stationary distribution. 
One concludes that a) this distribution is quite different from a uniform distribution, 
hence space charge has non-linear component; b) compared with a Gaussian distribution with the same rms parameters,
this distribution has a sharper peak, thinner waist and longer tail.  

\begin{figure}[ht!]
   \centering
 {\includegraphics[width=0.47\linewidth,trim=2.0cm 2cm 2.0cm 2cm]{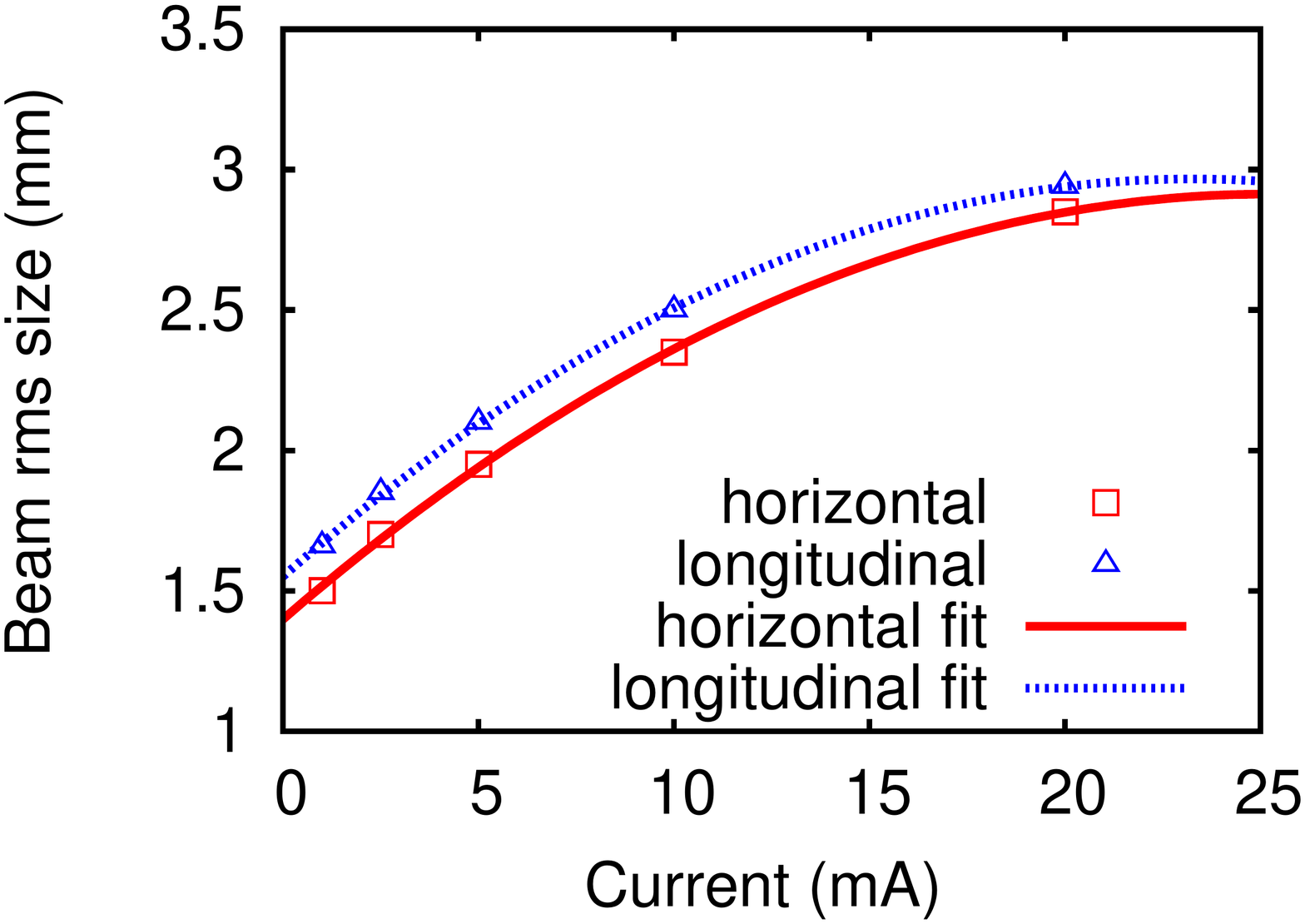}}
  {\includegraphics[width=0.47\linewidth,trim=2.0cm 2cm 2.0cm 2cm]{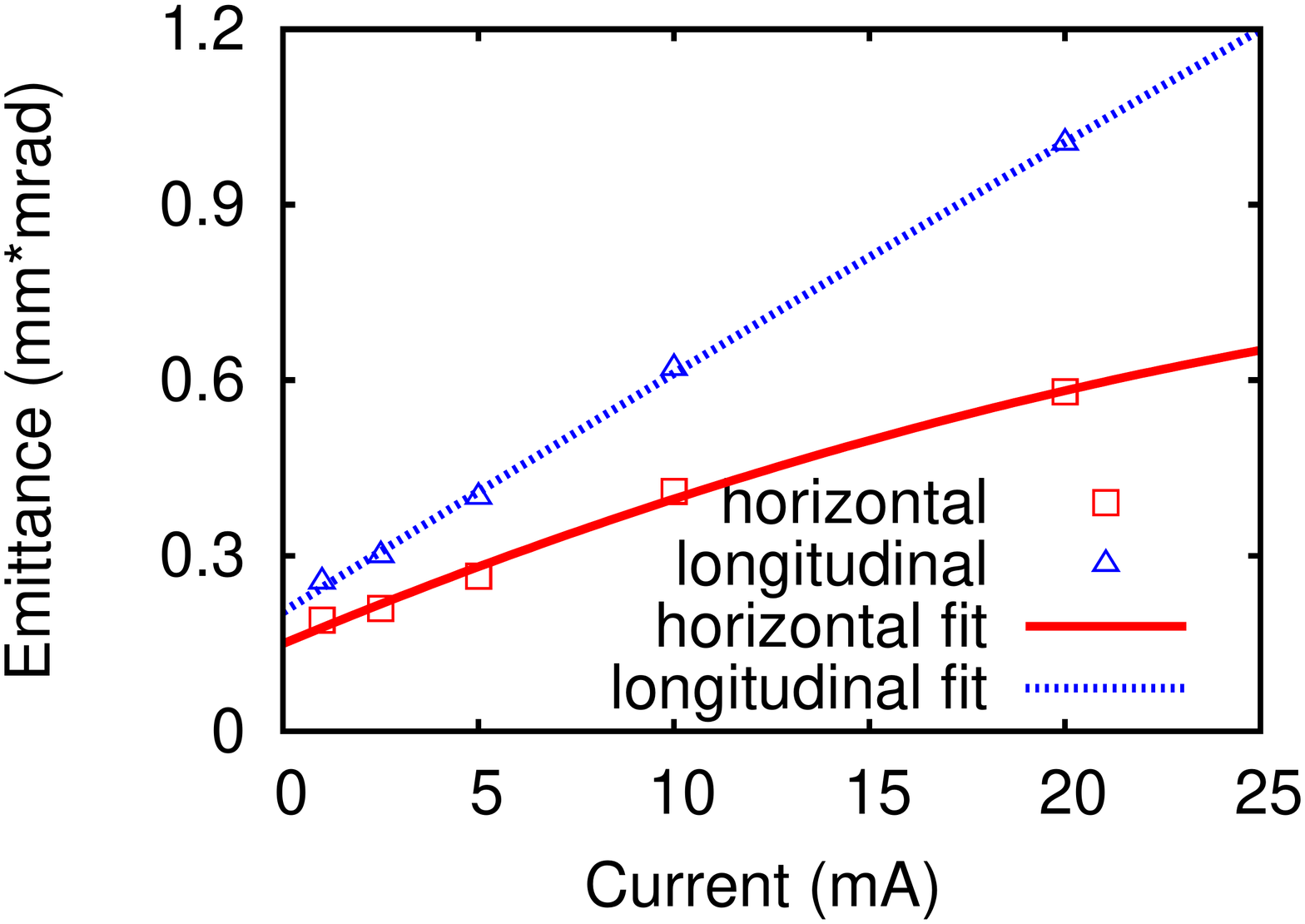}}
\caption{The stationary beam rms size (left) and the normalized rms emittance (right) vs beam current} % for the 0.5MeV beam
    \label{fig:d8}
  \end{figure}

Since the space charge force is proportional to the charge intensity in the bunch,
the size of the final matched beam is a function of the beam current.\  
Fig.\ \ref{fig:d8} compares the stationary beam size for the current from 1 mA to 20 mA that shows that both  the beam size and emittance of the formed stationary beam are gradually increase with increasing beam current.
The beam size increases along with current both  in the horizontal and longitudinal directions, however, the rate of growth is decreasing.
The beam  rms sizes can be well fitted by a quadratic polynomial 
\begin{equation} \label{eq:alpha}
\begin{array}{cc} 
	\sigma = aI^2+bI+c & (1 \mbox{ mA}\le I \le  20 \mbox{ mA}),
\end{array}
\end{equation}
where the values of $a$, $b$, $c$ are -0.0024, 0.12, 1.40  in the radial direction and  -0.0026, 0.12, 1.55 in the longitudinal direction respectively.
The beam emittance is more sensitive to the beam current than is the beam size. 
As the beam current increases  from 5 mA to 10 mA,  the horizontal and longitudinal emittances increase by approximately 50\%, 
but  the horizontal and longitudinal sizes increase only by approximately 20\%. 
% Fixme: no this is not correct 
%This can be explained from physics. The space charge force increase the momentum spread directly, 
%nevertheless the beam size is the time integration of the momentum spread. 
%Therefore, the beam size is less dependent on the beam current than the emittance. 

From the simulation, one concludes that the stationary matched distribution with a short length can be formed for a 5 mA beam current.\
In consequence, no flattop cavity is required, as is the case for the PSI Injector II cyclotron.\
Therefore, the four valleys are available for installing the accelerating cavities maximizing the energy gain per turn and
hence to obtain a sufficiently large turn separation at extraction to achieve the required low beam losses.

\subsection {Space charge effects during acceleration}
The Versatile Ion Source (VIS)  \cite{celona-1,maimone-1} will be used to provide 35 keV/amu \htp beam for the DIC cyclotron. 
The VIS is an off-resonance microwave discharge source operating at 2.45 GHz. 
It has demonstrated good CW proton currents ($>$35 mA) with a normalized emittance less than 0.2 $\pi$ mm-mrad 
and an extraction voltage which can be raised up to 70 kV. 

\begin{figure}[ht!]
\begin{center}
{\includegraphics[angle=-0, width=0.5\linewidth]{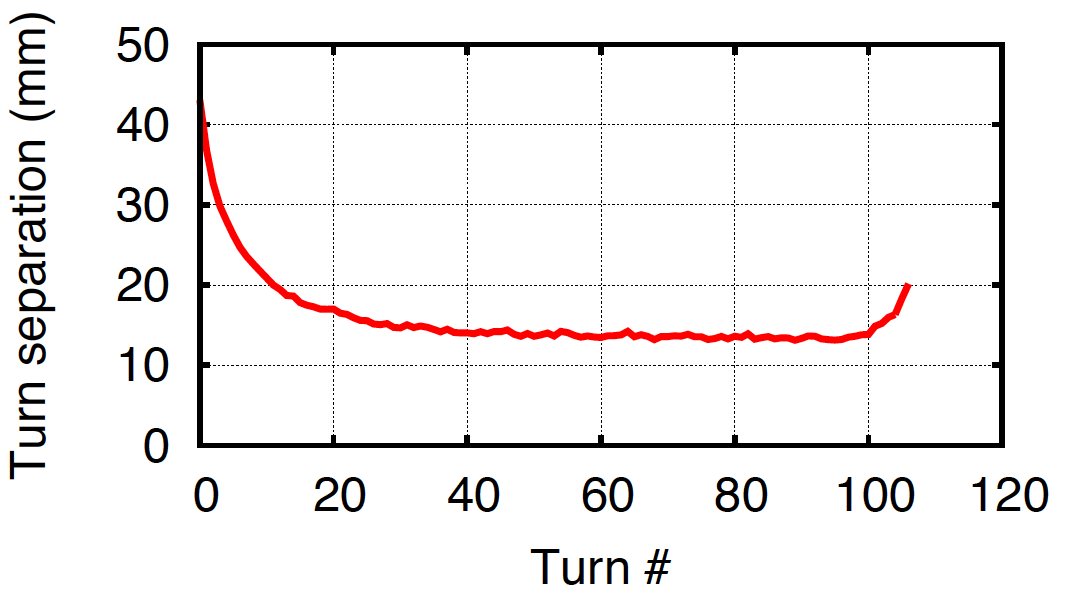}}
\end{center}
\caption{Turn separation of the central particle in the beam.
\label{fig:drdn}}
\end{figure}

The spiral inflector and central region of the DIC is in the design phase, and an experimental test of the central region is in preparation. 
The simulations of the DIC described in this paper start at the exit of the central region. 
Considering that both the space charge effects in the injection line and the transverse-longitudinal coupling motion in the spiral inflector inevitably  increase the emittance,
the initial normalized emittance at the exit of central region is set to three times larger than that of ion source, i.e., 0.6 $\pi$ mm-mrad.  %( $2\sigma$). 
The phase acceptance and initial energy spread are assumed to be $10^\circ-20^\circ$ and  0.4 \% respectively. %($2\sigma$).
In order to reduce the tail particles of the extracted beam,  four collimators are placed at around 1.9 MeV/amu to cut off about 10 \% of the halo particles. 
With the four collimators, the simulation is carried out for 1 mA, 5 mA  and 10 mA current respectively. 
To achieve the high statistics for halo particles,  $10^6$ macro particles are used the in the simulations
and $64^3$ rectangular grid are used for space charge calculation.
%Fig.~\ref{fig:lastturn} shows horizontal-longitudinal distribution of the last turn beam which will be extracted by the electrostatic deflector . 
\begin{figure}[ht!]
\begin{center}
{\includegraphics[angle=-0, width=0.5\linewidth]{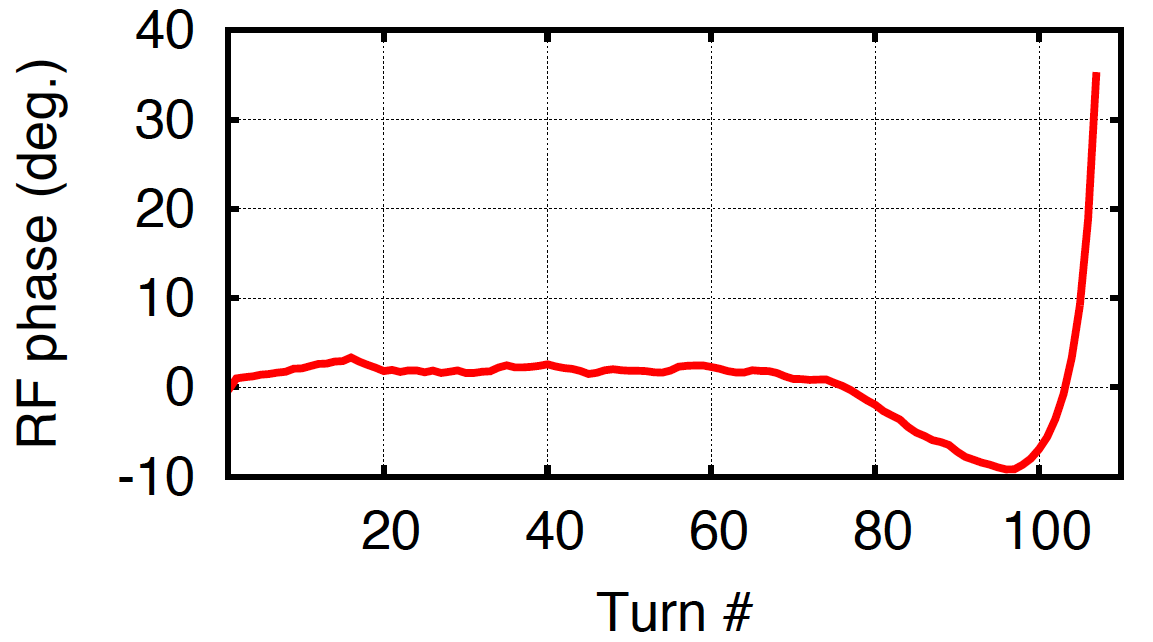}}
\end{center}
\caption{RF phase slipping of the central particle in the beam
\label{fig:rfphase}}
\end{figure}
The \htp beam needs 107 turns to reach the final energy. The phase slipping during the acceleration is plotted in  Fig.\ \ref{fig:rfphase},
showing the phase slipping is within $\pm10^\circ$ in the first 100 turns. During the final turns it increased by $45^\circ$, 
caused by the a sharp field drop, needed to increase the turn separation by passing the $\nu_r$=1 resonance.  
The total turn separation consists of two parts:
\begin{equation}\label{eq:dR}
\frac{dR}{dn} = \frac{dR}{dn}\mbox{(accel)}+\frac{dR}{dn}\mbox{(precession).}
\end{equation} 
The contribution of acceleration can be expressed by
\begin{equation}\label{eq:dRaccel}
\frac{dR}{dn}\mbox{(accel)} = R \times\frac{\Delta E } {E} \times\frac{\gamma}{\gamma+1}\times\frac{1}{\nu^2_r},
\end{equation} 
where $R$ denoting the radius, $\Delta E$ is the energy gain per turn, $E$ is the total energy of the
particles, and $\nu_r$ is the radial focusing frequency. 
The maximum turn separation of precession motion can be expressed by
\begin{equation} \label{eq:dRprecss}
\frac{dR}{dn}\mbox{(precession)}=2\times x_c\times \sin{\pi(1-\nu_r)},
\end{equation}
where $x_c$ is a coherent betatron oscillation amplitude of beam. 
Figure \ref{fig:drdn} shows the turn separation obtained by particle tracking.
The separation of the last turn is $\sim$ 20 mm, in very good agreement  with Eq.\ (\ref{eq:dR}),
i.e. the contribution of acceleration and precession are 13 mm and 7 mm respectively. 

\begin{figure}[ht!]
\centering
{\includegraphics[width=0.45\linewidth,trim=2.0cm 3cm 4cm 2cm, keepaspectratio=false]{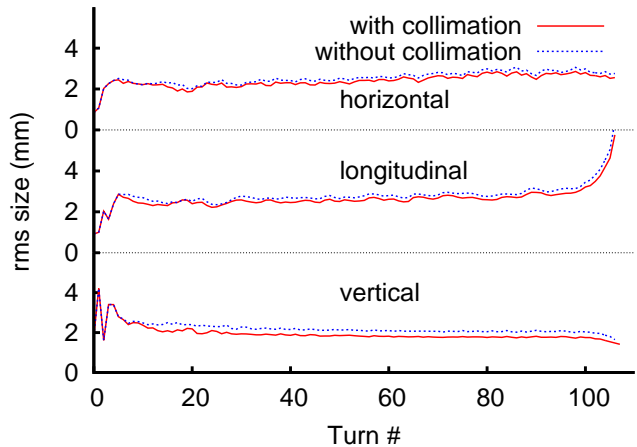}
\caption{The rms size snapshot at $0^\circ$ azimuth the during acceleration with and without collimation for a 5 mA beam.}
\label{fig:DICacc}}
\end{figure}

Figure \ref{fig:DICacc} shows the rms beam size during  acceleration.  
Again the beam sizes changes rapidly during the first five turns. This time the mismatch is mainly caused by the acceleration, 
which was not taken into account when finding the matched distribution of the coasting beam.
After an particle repartitioning process driven by space charge force, 
the horizontal and longitudinal dimensions converge to nearly same size and only slowly increase at higher energies. 
A direct consequence is the reduction of the longitudinal phase width and, as a result,  the suppression of the energy spread.
%yang: the referee may ask why the longitudinal  beam size blow up at extraction.
During the last few turns the longitudinal beam size increases rapidly because the longitudinal space charge focusing is destroyed by the phase slipping. 
However, the radial beam size is unaffected by the phase slipping, a point that  is important for extraction.  
The vertical beam size gradually decreases because of the increase of external focusing.
The effect of the four collimators on the rms beam size is small as expected.
% jianjun, stop here on the airplane

\subsection{Single-turn extraction from the DIC} 
\begin{figure}[ht!]
\begin{center}
{\includegraphics[width=0.48\linewidth, trim=2cm 4cm 0cm 0cm, keepaspectratio=false]{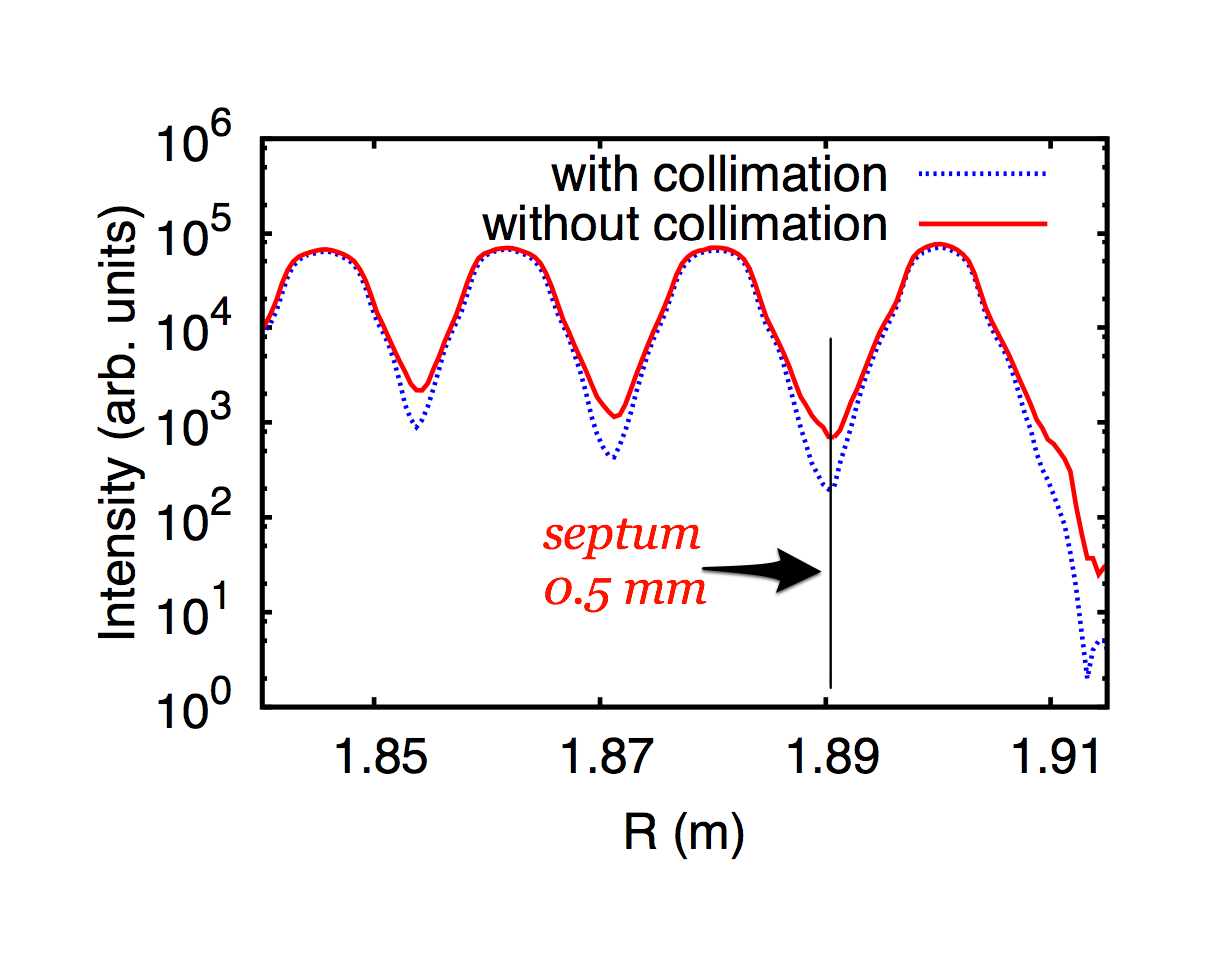}}
{\includegraphics[width=0.48\linewidth,  trim=2cm 4cm 0cm 0cm, keepaspectratio=false]{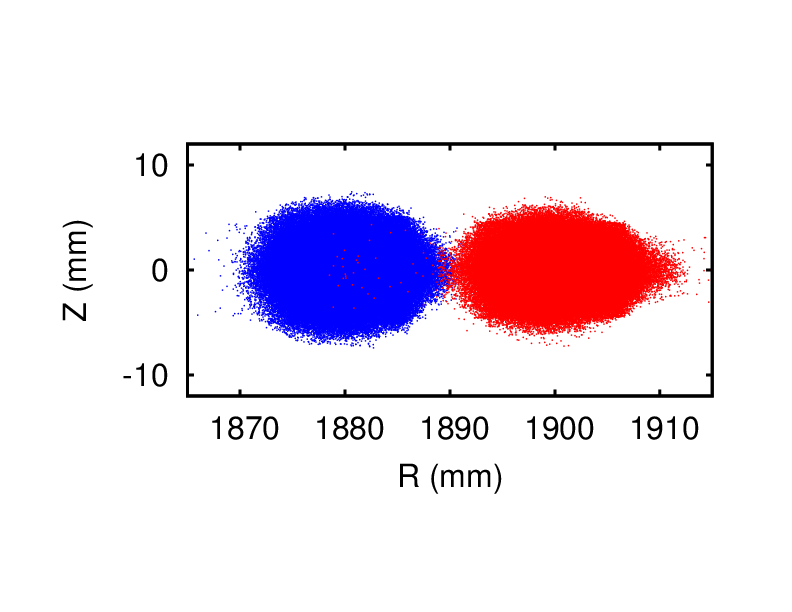}}
\end{center}
\caption{The radial profile of the last 4 turns at the center on the valley  for 5mA current (left)
 and the the r-z projection of last 2 turns particle distribution for the case with collimation (right). The total macro particle number in a bunch is $10^6$.
\label{fig:5mA10deg}}
\end{figure}

 In OPAL, a radial profile probe element enables one to compute the radial beam profile at the location of  the electrostatic deflector. 
 Figure \ref{fig:5mA10deg} (left) shows the radial beam profile at the position of the electrostatic deflector during the last four turns.
One sees that the implemented collimation scheme reduces the charge intensity at the deflector position by a factor of four
--- a large improvement with respect to controlled beam losses at the deflector. 
 Figure\ \ref{fig:5mA10deg} (right) shows the radial-vertical particle distribution  at the deflector position during the last two turns.
A very conservative septum thickness of 0.5 mm keeps the beam loss at extraction to less than 120 W. 
In the DAE$\delta$ALUS  experiment, the DIC will work at the a duty cycle of 20\%, hence the beam power at the deflector is less than 24 W.
The Isotope Decay-at-Rest (IsoDAR) \cite{isodar} experiment is an intermediate step of DAE$\delta$ALUS, in which the DIC will be used as a low energy high intensity driver. The IsoDAR electron antineutrino source requires the DIC to work at a duty cycle of 90\% ,
in which case the beam power will be less than 108 W. Based on the experience with the PSI Injector II cyclotron,
we conclude that the most crucial part of the DIC, the extraction is feasible. 

\begin{figure}[ht!]
\begin{center}
{\includegraphics[width=0.45\linewidth, trim=0cm 3cm 0cm 2cm, keepaspectratio=false]{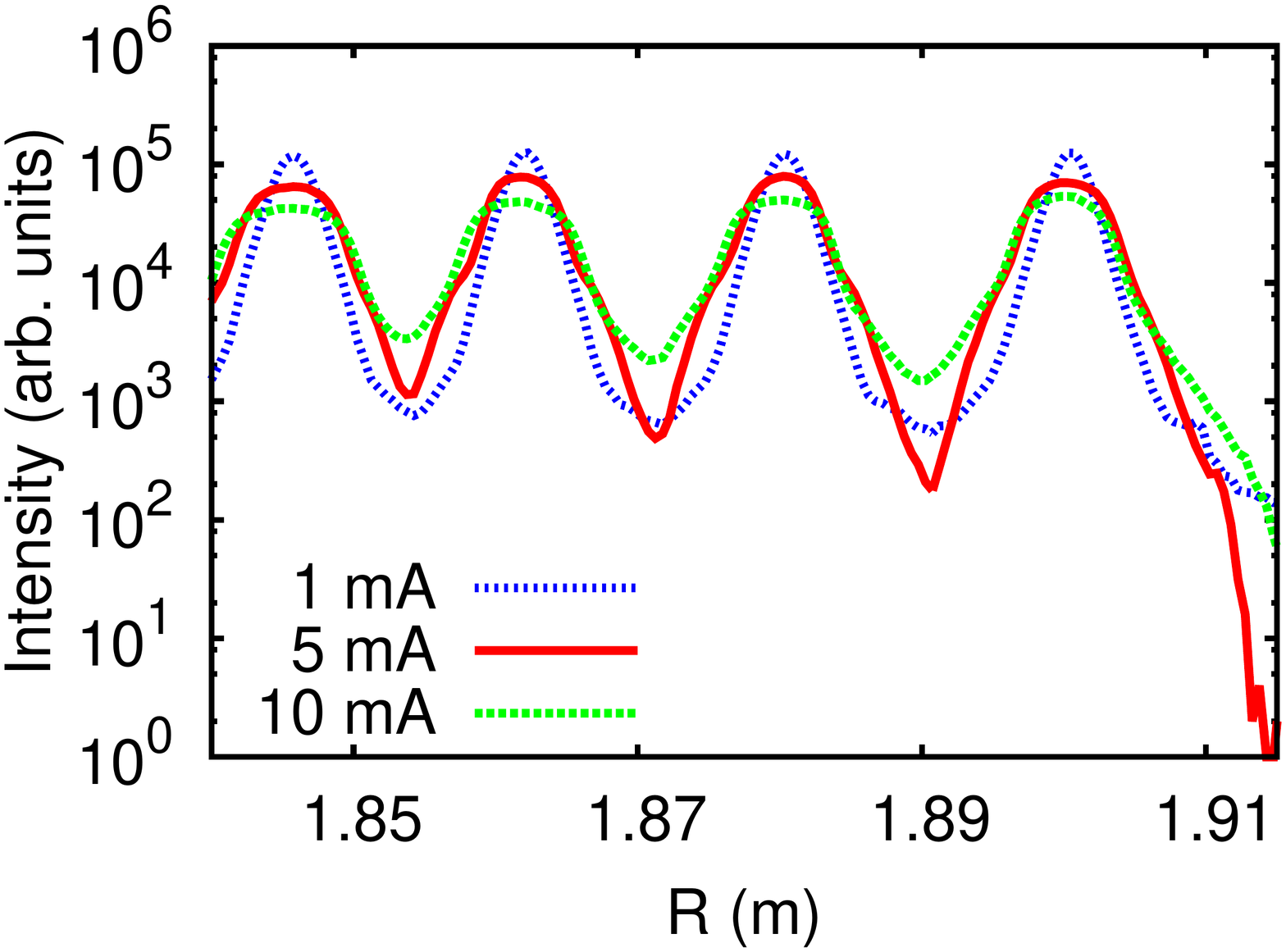}}
{\includegraphics[width=0.45\linewidth, trim=0cm 3cm 0cm 2cm, keepaspectratio=false]{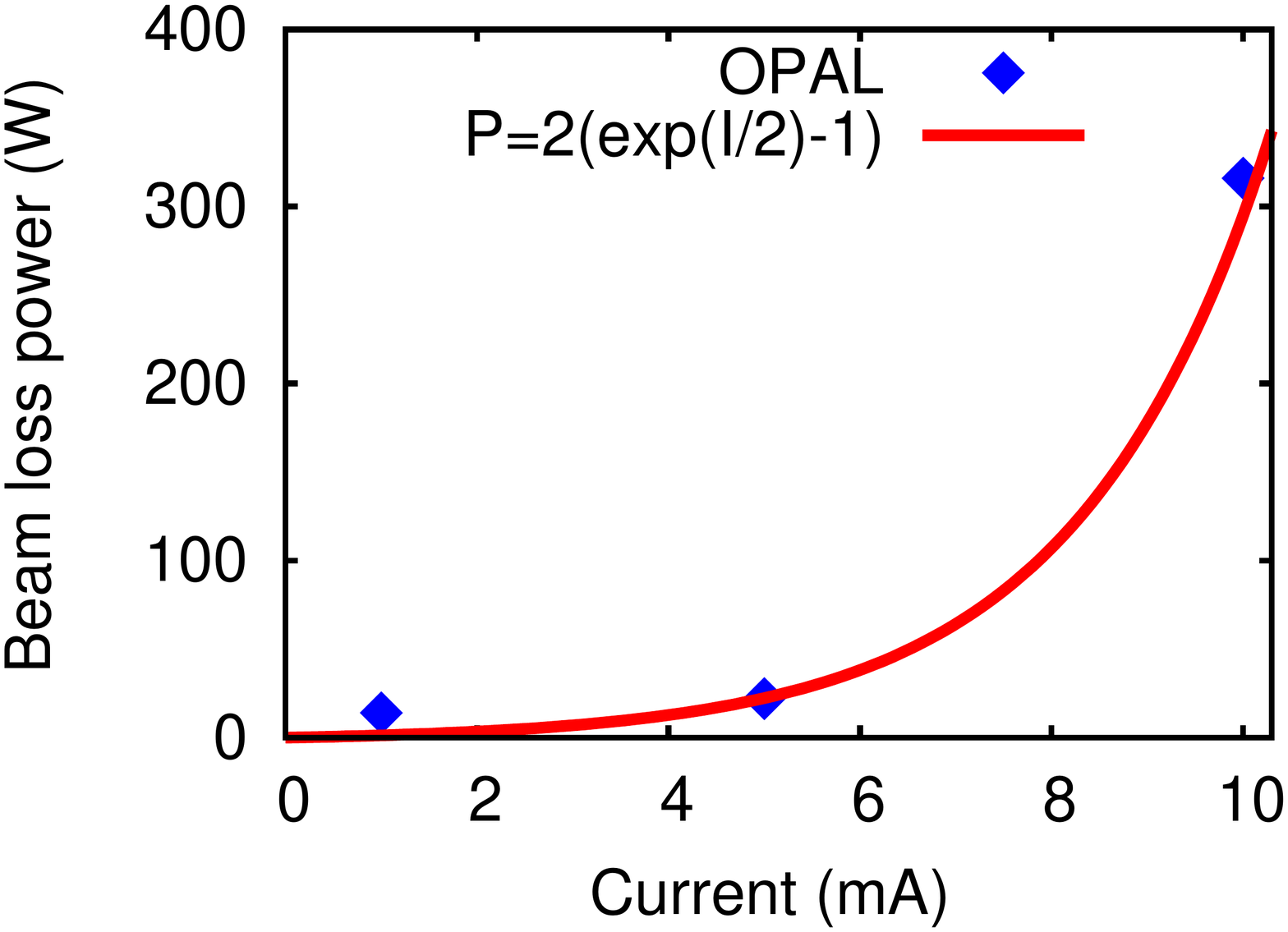}}
\end{center}
\caption{Left: the radial profile of the last 4 turns at the center on the valley for the different beam current with the initial phase width of $20^\circ$;
right: the resultant beam loss power on septum at 20\% duty cycle and the exponential fitting curve. The total macro particle number in a bunch is $10^6$.
\label{fig:profileI}}
\end{figure}

To evaluate the sensitivity of the beam losses to variation of injection conditions, 
we simulated scenarios with different phase acceptances in the central region and different beam currents. 
Figure \ref{fig:profileI} (left) shows the radial beam profile during the last four turns for 1 mA, 
5 mA and 10 mA average beam current with initially $20^\circ$ phase width. 
The radial  beam sizes broaden with increased current thereby increasing the beam loss on the septum.
An interesting phenomenon is that at the deflector septum position, for the 1 mA beam has more halo particles (i.e., the higher relative beam loss)
than that of the 5 mA case.  
This behaviour is attributed to the fact, that the 1 mA beam is at the transition between the emittance-dominated and space-charge-dominated.
Therefore beam distribution is more easily compromised by the phase slipping.
Fig.\,\ref{fig:profileI} (right) shows the exponential increase of  the power of beam loss on the septum versus beam current, which is well fitted by the exponential function
\begin{equation}\label{eq:beamloss}
 P=2\left(\exp(I/2)-1\right).  
\end{equation} 

Table~\ref{tab:beamloss} summarizes the beam loss for different beam currents and injection phases. 
At the level of 5 mA, the beam loss on the septum is not sensitive to the injection phase width
thereby allowing the flexibility for the central region design.
When the current increases to 10 mA, the beam loss on septum increases tenfold to a level which exceeds the tolerable level of loss. 

\begin{table}
\caption {\label {tab:beamloss}The beam loss power on the septum for 20\% duty cycle (uints: W).}
  \begin{tabular}{cccc}
  \hline \hline   
	injection phase width & 1 mA & 5 mA &10 mA \\ 
	\hline
	$10^\circ$  & 6 &24  & 240 \\
	$20^\circ$  & 14 &22  & 316 \\
	\hline \hline
  	\end{tabular}	 	
\end{table}

\section{ Space charge in the DSRC cyclotron\label{DSRC}}
In the DSRC, the beam loss at extraction is relaxed by using stripping extraction which has an efficiency close to 100\%, 
even if the beam orbits at the extraction radius are not well separated.
Production of neutral hydrogen $H_0$ is a significant problem in stripping $H^-$,
%At SNS they observe about 3\% of their beam emerging from the stripper as $H_0$.
but for the \htp beam the fraction of $H_0$  is expected to be substantially lower, as there is only one electron to remove, 
and this electron is much less tightly bound (2.7 eV) than an electron in neutral hydrogen (13 eV). 
The main challenge comes from two other aspects:
a) the vertical focusing at higher energies, which is usually weak in a superconducting cyclotron;
b) the relatively large energy spread of the resulting proton beam, due to the nature of multi-turn extraction.
The following section investigates these two questions.
%which will enlarge proton beam envelops along the extraction path.

\subsection {Space charge effects during acceleration}

\begin{figure}[ht!]
\begin{center}
    {\includegraphics[width=0.49\linewidth, trim=0.5cm 3.0cm 0.0cm 0.0cm]{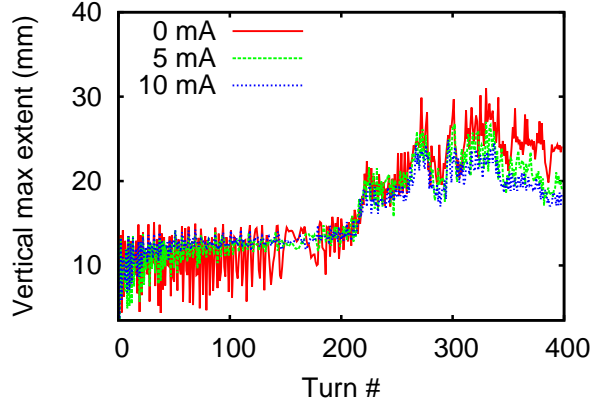}}  
\end{center}
\caption{The maximum beam extent in the vertical direction in the acceleration
\label{fig:zprofile}}
\end{figure}

\begin{figure}[ht!]
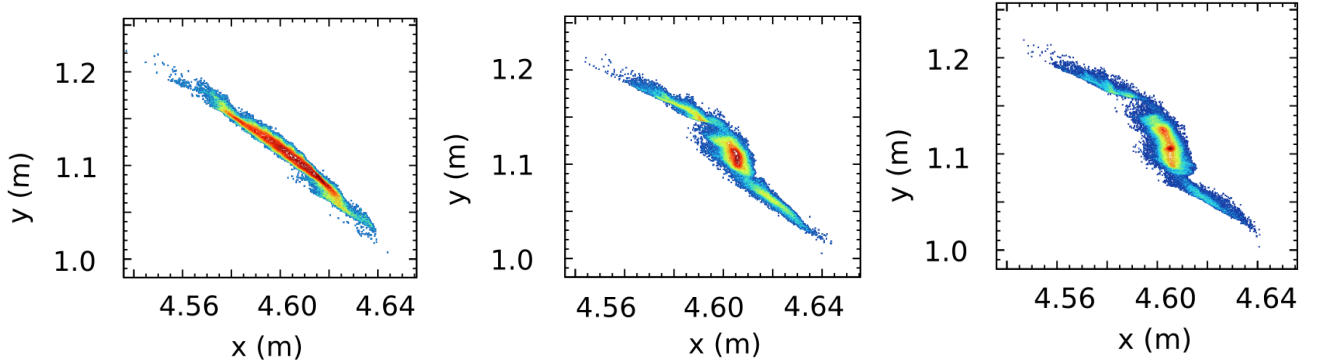

\begin{center}
 {\includegraphics[width=0.32\linewidth]{0mA32-x-y-step-394}}     
 {\includegraphics[width=0.32\linewidth]{5mA32-x-y-step-394}}     
 {\includegraphics[width=0.32\linewidth]{10mA32-x-y-step-394}}     
\end{center}
\caption{Top view of the final turn for 1 mA (left), 5 mA (middle) and 10 mA (right) beam current
\label{fig:topview}}
\end{figure}

Since the layout of the transfer line between the DIC and the DSRC is not yet fixed,
in the simulation we assume a typical gaussian distribution with conservative values for the normalized rms emittances of 0.9 $\pi$mm-mrad 
in both the transverse directions, full phase width of $\pm10^\circ$ and energy spread of 0.1\%. 
Four single-gap cavities with a peak voltage of 1 MV accelerates the \htp beam in 401 turns up to an energy 800 MeV/amu.
Figure \ref{fig:zprofile} shows the maximum vertical  extent of the beam during acceleration for different current scenarios. 
Although the vertical extent of the beam is insensitive to space charge, but the beam extents gradually to 30 mm (full width) because of the weak external focusing, 
starting around turn 220. As the height of the magnet gap is 80 mm, no significant beam loss is expected. 
Figure \ref{fig:topview} shows the longitudinal-radial configuration space of the final turn at extraction; 
the beam extends to $\sim$20 cm for different intensities. A stationary, compact shape will not develop, 
because the strength of the space charge force scales with $1/\gamma^2$. 
Hence the needed force for the vortex motion is too small. For the cases of 5 mA and 10 mA, 
the beam splits into three sub-bunches longitudinally that exemplify the longitudinal instability introduced in Ref.\ \cite{Pozdeyev:2, Bi:1}.

For high intensity cyclotrons with small turn separation, 
single bunch space charge effects are not the only contribution to beam dynamics. Along with the increase of beam current, 
the mutual interaction of neighboring bunches in the radial direction becomes ever more important.  
In the DSRC, the turn separation at injection is 25 mm  and gradually decreases to 3 mm at extraction.
Therefore, the interaction of neighboring bunches is of greatest importance at large radii where the bunches overlap and where the extraction foil is located.  
We have used  the ``Start-to-Stop" model \cite{Yang:2} which is implemented in OPAL code to study these effects in detail.

\begin{figure}[ht!]
\begin{center}
      {\includegraphics[width=0.8\linewidth, trim=0.0cm 0.0cm 0.0cm 0cm]{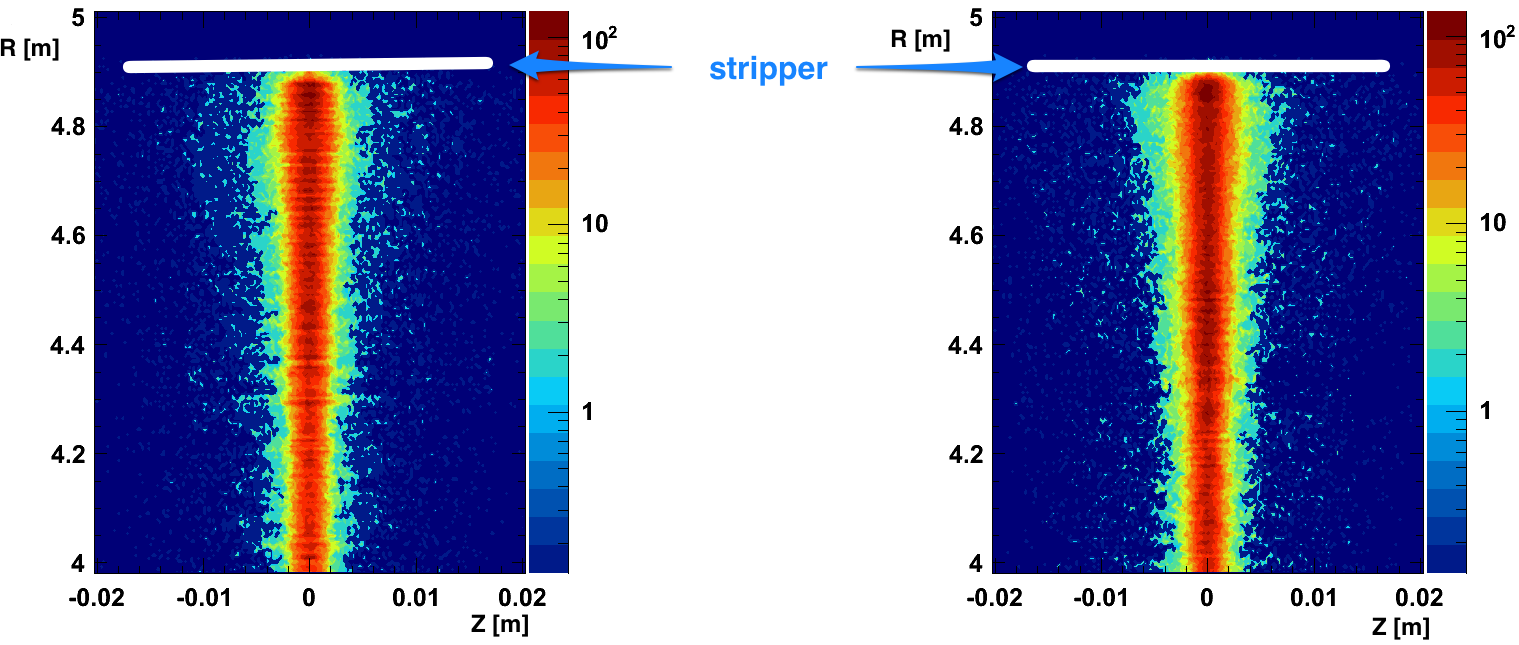}} 
\end{center}
\caption{The projection of beam intensity onto the vertical \& radial plane. 
\label{fig:MB}}
\end{figure}

The simulation starts with a single bunch and automatically transfers to the multi-bunch tracking mode, when radially neighboring bunch effects become important. 
Figure\ \ref{fig:MB} shows  the projection of phase space onto the vertical (Z) - radial (R) plane for a 0 mA (left) and a 5 mA (right) case.
Beam halos extends vertically to $\pm$20 mm which still clears the magnet sector and vacuum chamber.
We conclude from the simulation that in the DSRC space charge and  neighboring bunch effects introduce vertical beam halo, 
but the influence of those effects is at an acceptable value.

\subsection{\htp stripping}
\begin{figure}[ht!]
\begin{center}
 {\includegraphics[width=0.32\linewidth,trim=2.0cm 2cm 1cm 2cm, keepaspectratio=false]{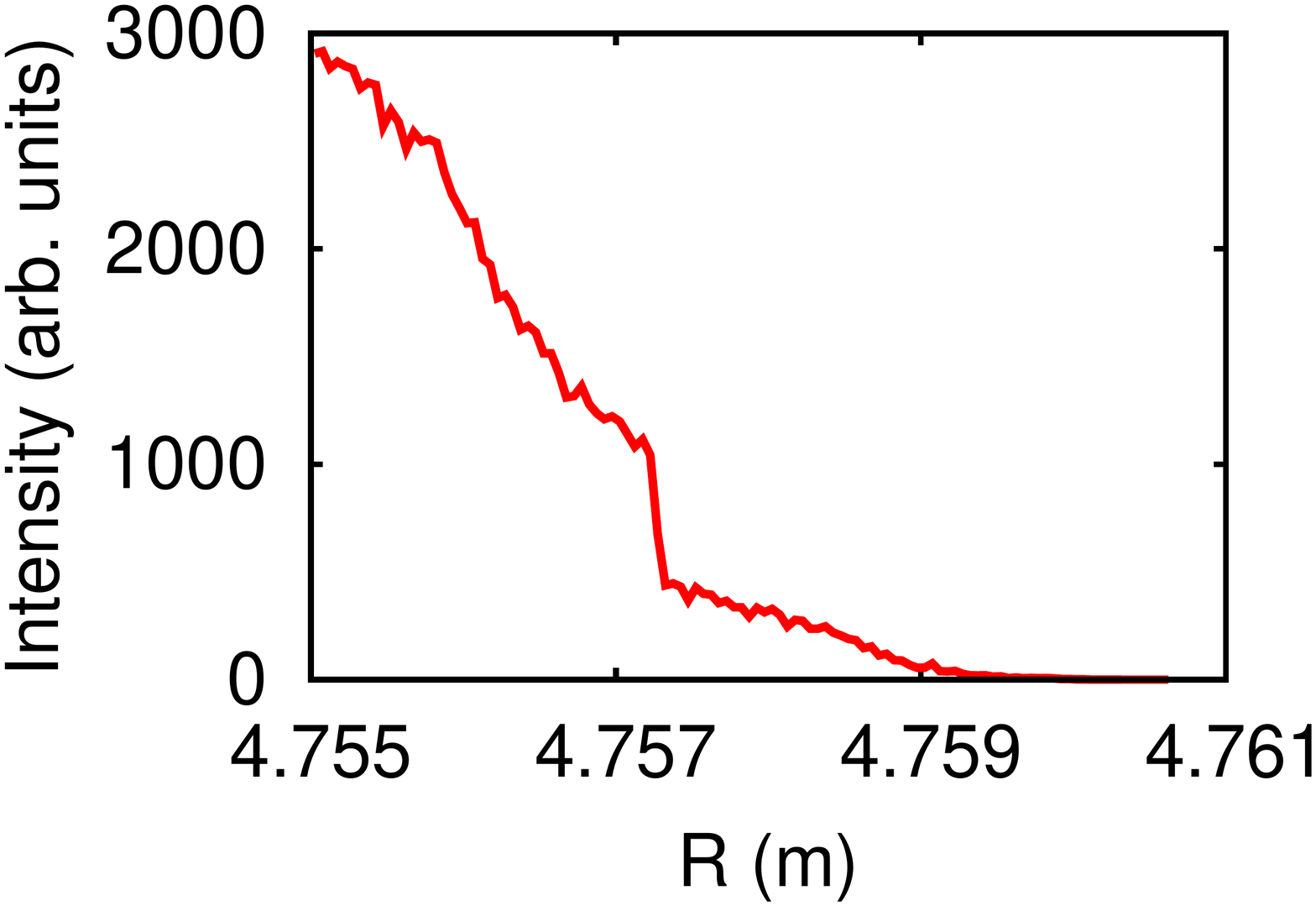}}
 {\includegraphics[width=0.32\linewidth,trim=2.0cm 2cm 1cm 2cm, keepaspectratio=false]{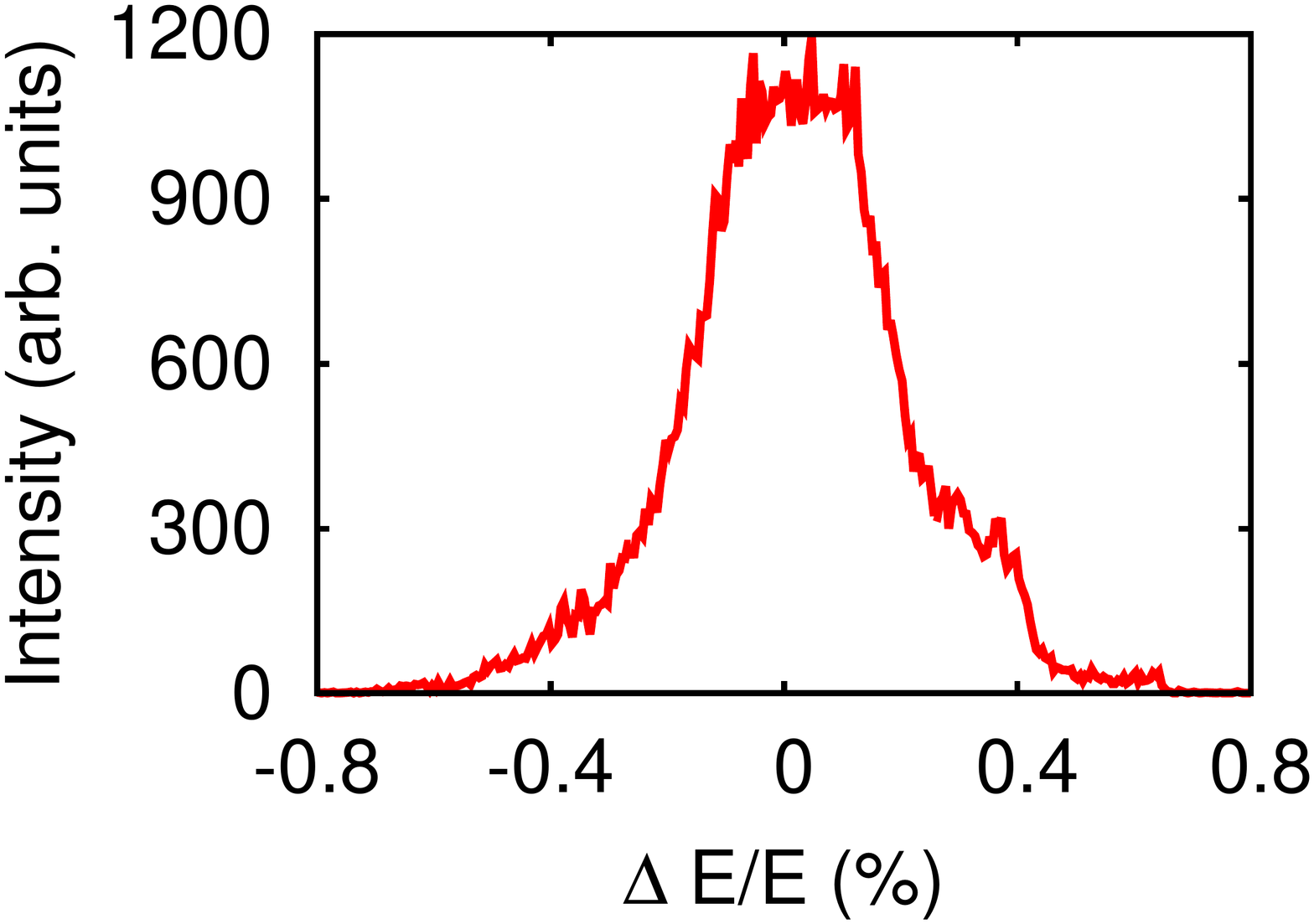}}
 {\includegraphics[width=0.32\linewidth,trim=2.0cm 2cm 1cm 2cm, keepaspectratio=false]{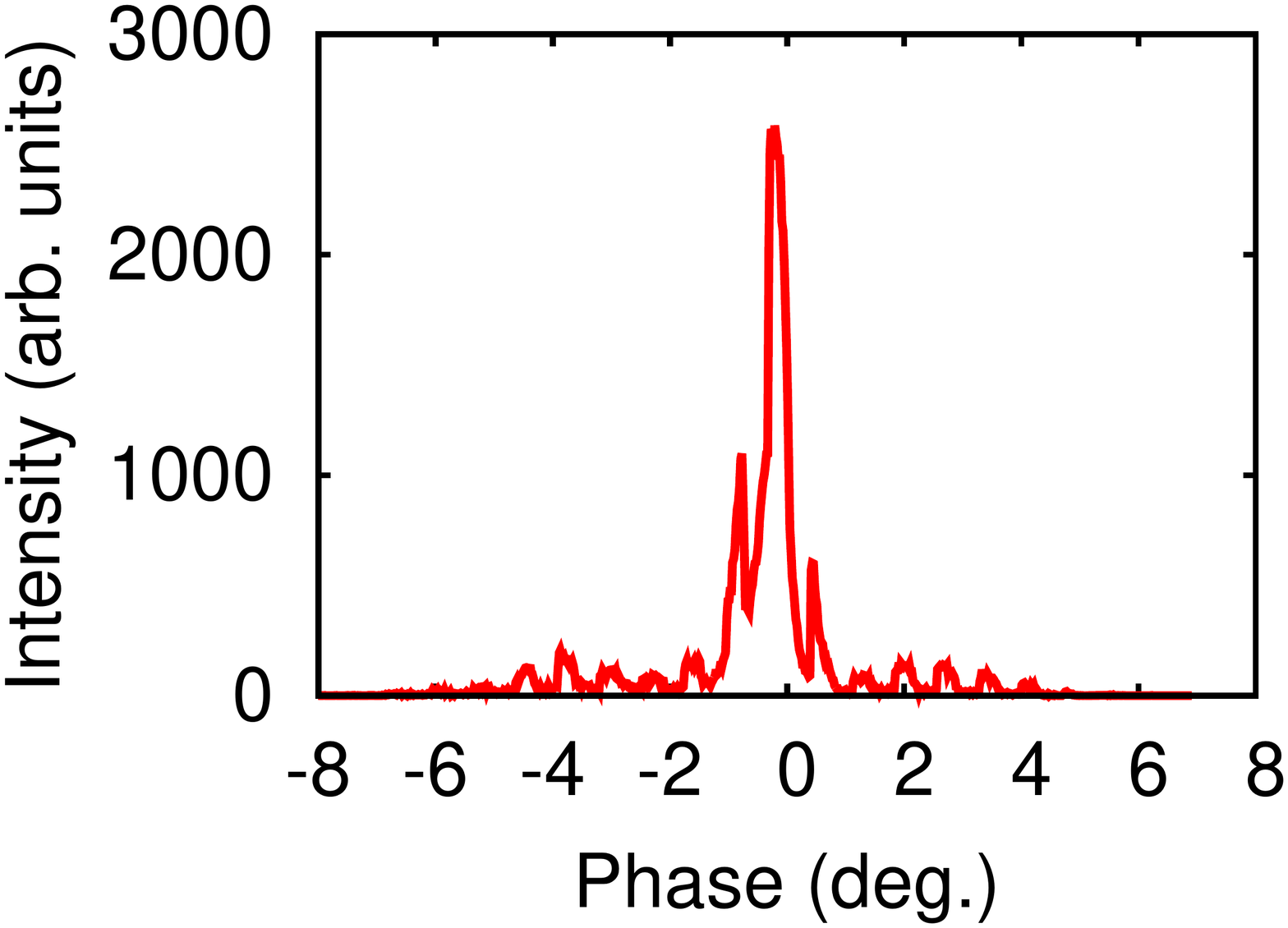}}
\end{center}
\caption{Histograms of the proton distribution on the foil.
\label{fig:stripper}}
\end{figure}

In the current design of the DSRC, the two strippers are located at 35$^\circ$ and 215$^\circ$ azimuth respectively (as are shown in Fig.\ \ref{fig:porbit}).
The stripper-element in OPAL samples the 6D-particle-distribution, and converts one \htp into two protons. 
OPAL contains no physics processes for particle foil interaction and the creation of associated electrons, however 
dedicated programs are available for this set of problems.
Figure \ref{fig:stripper} shows histograms of different beam quantities at the stripper foil. For this particular data set, 
the last 15 turns are contribute, exemplifying the need of a truly multi-bunch simulation.

At extraction the full energy spread is  about 1.2\%, which is 12 times the energy spread of at injection. 
In the  longitudinal direction, we see a long tail and the full phase width is  $10^\circ$.
The normalized rms emittances are $\varepsilon_{r}$=1.1  $\pi$mm-mrad and  $\varepsilon_{z}$=1.4  $\pi$mm-mrad,  
as compared with the injection emittance of  0.9  $\pi$mm-mrad for both planes.

The particle distribution on the stripper foil is taken as the initial distribution for the calculation of the foil temperature and lifetime calculation. 
These results will be reported in a subsequent paper.

\subsection {Proton beam extraction}
\begin{figure}[ht!]
\begin{center}
 {\includegraphics[width=0.5\linewidth,,trim=4.0cm 4cm 4cm 1cm]{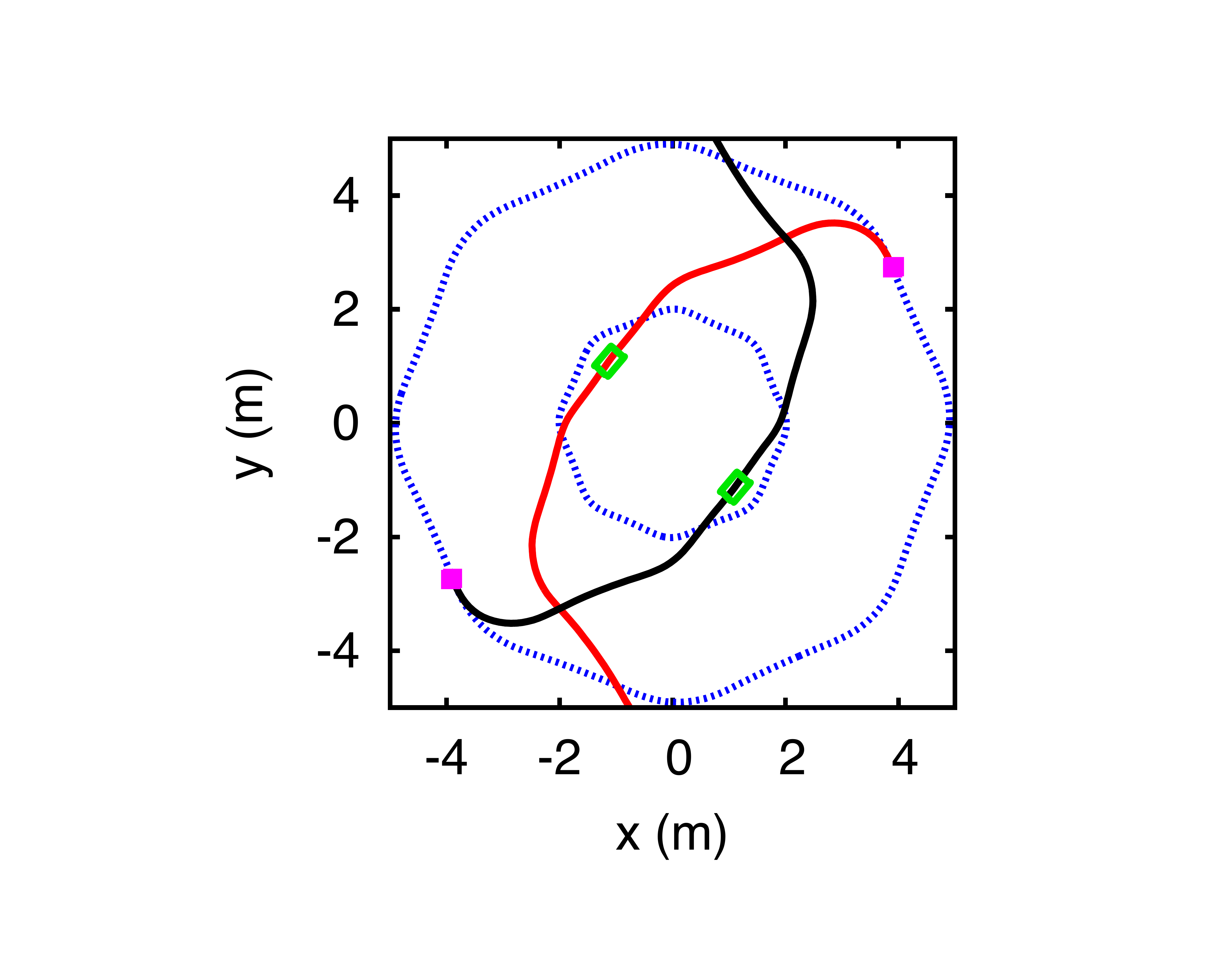}} 
\end{center}
\caption{The two proton extraction trajectories(red and black) and the strippers at 35$^\circ$ and  215$^\circ$ (magenta point), 
together with the first and last turn of the  reference \htp acceleration orbit (blue) and the shape of the extra dipole filed (green rectangle).
Other intermediate acceleration turns are not shown for clearance.
\label{fig:porbit}}
\end{figure}

Since the extracted protons have only half of the $H_2^+$ magnetic rigidity, 
the beam will be bend first towards the center of the machine and then towards the exit port.
In this process the particles will experience complicated electrical-magnetic fields, crossing fringe fields and rf-cavities. 
Figure\ \ref{fig:porbit} shows the reference proton trajectory, together with two selected \htp orbits for illustration purposes. 
The simulation shows that the vertical beam size extends into the hill gap in the region of the exit port. 
In order to compensate for this, an extra dipole field with 1.6 kG/cm is applied at the inner free space to strengthen the vertical focusing, 
as is shown in Fig.\,\ref{fig:porbit}.

\begin{figure}[ht!]
\begin{center}
 {\includegraphics[width=0.6\linewidth,trim=0.0cm 4cm 0cm 4cm]{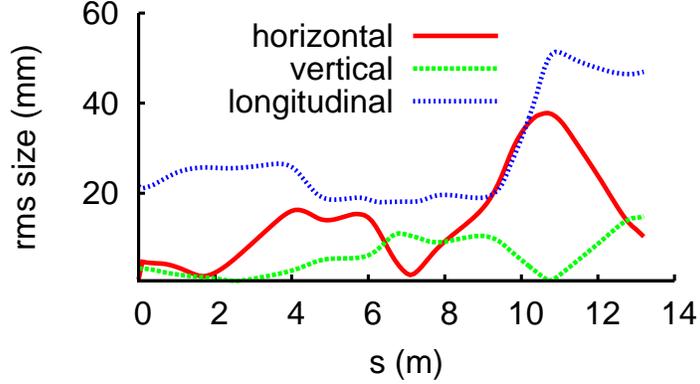}}
\end{center}
\caption{The proton beam envelops along the extraction trajectory.
\label{fig:penvelop}}
\end{figure}

The proton beam is tracked from the particle distribution on the stripper to the exit port and the extraction path. 
Figure \ref{fig:penvelop} shows the beam envelope along the extraction trajectory.
It can be seen that the beam is well-focused vertically with the help of the extra dipole field. 
The energy spread enlarges the longitudinal beam size, significantly, while the dispersion increases the horizontal beam envelop to 40 mm. 
We expect that the aperture of the extraction line towards the target will accept this beam without significant difficulties.

\section{Conclusion and discussion}

Given the tremendous impact of space charge on the beam's behavior in high intensity cyclotrons,
precise beam dynamics simulation has been included at the very beginning of the conceptual design of  the DAE$\delta$ALUS  cyclotrons. 
Based on the space charge study reported here, the current version of the cyclotron structure is well optimized with respect to the beam loss.

The simulation shows that beam transport in the DIC cyclotron is space-charge-dominated allowing the formation the stationary matched distribution with short length.
Therefore, no flattop cavity is required and the four valleys are available for installing the accelerating cavities.
The \htp beam can be extracted with beam loss on septum less than 150 W for 100\% duty cycle, 
which will not result in serious problems at extraction.

Furthermore, in the DSRC the important beam properties at the stripper are dominated by the initial conditions. 
Space charge, and radially neighboring bunches, introduce vertical beam halo, but the influence is at acceptable levels.
The multi-turn stripping scheme relaxes the constraint of beam loss at extraction.
However, with the side effects of a larger emittance and larger energy spread of the extracted beam.
Fortunately, the proton beam tracking along the extraction path shows the  beam is well focused and therefore no significant beam loss is expected. 

Furthermore, other sources of beam  losses which are not discussed in this paper. These phenomena include Lorentz dissociation of unquenched vibrational states of \htp particles, collisions with residual gas and incomplete foil stripping. Studies of these aspects will follow in future papers.

\section{Acknowledgment}
The authors would like to extend sincere thanks to A. Calanna and D. Campo for providing the field maps of the DSRC and DIC cyclotron.
We also feel indebted to M. Toggweiler, C. Kraus and Y. Ineichen, H. Zhang, R. D$\ddot{o}$lling and 
C. Baumgarten for helpful discussions. The majority of computations have been performed on the local PSI computing resources FELSIM and Merlin4 and 
part of the data are analyzed using the visualization tool H5root.
\bibliography{Refbase}

\begin{thebibliography}{24}
\expandafter\ifx\csname natexlab\endcsname\relax\def\natexlab#1{#1}\fi
\expandafter\ifx\csname bibnamefont\endcsname\relax
  \def\bibnamefont#1{#1}\fi
\expandafter\ifx\csname bibfnamefont\endcsname\relax
  \def\bibfnamefont#1{#1}\fi
\expandafter\ifx\csname citenamefont\endcsname\relax
  \def\citenamefont#1{#1}\fi
\expandafter\ifx\csname url\endcsname\relax
  \def\url#1{\texttt{#1}}\fi
\expandafter\ifx\csname urlprefix\endcsname\relax\def\urlprefix{URL }\fi
\providecommand{\bibinfo}[2]{#2}
\providecommand{\eprint}[2][]{\url{#2}}

\bibitem[{\citenamefont{Conrad and Shaevitz}(2010)}]{Conrad:1}
\bibinfo{author}{\bibfnamefont{J.~M.} \bibnamefont{Conrad}} \bibnamefont{and}
  \bibinfo{author}{\bibfnamefont{M.~H.} \bibnamefont{Shaevitz}},
  \bibinfo{journal}{Phys. Rev. Lett.} \textbf{\bibinfo{volume}{104}},
  \bibinfo{pages}{141802} (\bibinfo{year}{2010}).

\bibitem[{\citenamefont{{J.~Alonso, F.~T.~Avignone, W.~A.~Barletta, R.~Barlow,
  H.~T.~Baumgartner, A.~Bernstein, E.~Blucher and L.~Bugel {\it et
  al.}}}(2010)}]{Alonso:2010fs}
\bibinfo{author}{\bibnamefont{{J.~Alonso, F.~T.~Avignone, W.~A.~Barletta,
  R.~Barlow, H.~T.~Baumgartner, A.~Bernstein, E.~Blucher and L.~Bugel {\it et
  al.}}}} (\bibinfo{year}{2010}).

\bibitem[{\citenamefont{{M. Abs, {\it et al.}}}(2012)}]{EricePaper}
\bibinfo{author}{\bibnamefont{{M. Abs, {\it et al.}}}},
  \bibinfo{journal}{{arXiv:1207.4895}}  (\bibinfo{year}{2012}).

\bibitem[{\citenamefont{Calabretta et~al.}(2010)\citenamefont{Calabretta,
  Maggiore, Piazza, and Rifuggiato}}]{Luciano:1}
\bibinfo{author}{\bibfnamefont{L.}~\bibnamefont{Calabretta}},
  \bibinfo{author}{\bibfnamefont{M.}~\bibnamefont{Maggiore}},
  \bibinfo{author}{\bibfnamefont{L.~A.~C.} \bibnamefont{Piazza}},
  \bibnamefont{and}
  \bibinfo{author}{\bibfnamefont{D.}~\bibnamefont{Rifuggiato}}, in
  \emph{\bibinfo{booktitle}{Proc. 19th Int. Conf. on Cyclotrons and their
  Applications}} (\bibinfo{address}{Lanzhou}, \bibinfo{year}{2010}), p.
  \bibinfo{pages}{TUA1CIO01}.

\bibitem[{\citenamefont{{L. Calabretta, {\it et al.}}}(2011)}]{Luciano:2}
\bibinfo{author}{\bibnamefont{{L. Calabretta, {\it et al.}}}},
  \bibinfo{journal}{arXiv:1107.0652v1 [physics.acc-ph]}
  (\bibinfo{year}{2011}).

\bibitem[{\citenamefont{Reiser}(2008)}]{reiser2008theory}
\bibinfo{author}{\bibfnamefont{M.}~\bibnamefont{Reiser}},
  \emph{\bibinfo{title}{Theory and design of charged particle beams}}, Wiley
  Series in Beam Physics and Accelerator Technology
  (\bibinfo{publisher}{Wiley-VCH}, \bibinfo{year}{2008}), ISBN
  \bibinfo{isbn}{9783527407415},
  \urlprefix\url{http://books.google.ch/books?id=eegK9Mqgpi4C}.

\bibitem[{IBA(2012)}]{IBA}
\bibinfo{type}{Tech. Rep.},
  \bibinfo{institution}{http://www.iba-cyclotron-solutions.com/products-cyclo/cyclone-30}
  (\bibinfo{year}{2012}).

\bibitem[{cyc(2012)}]{cyclotron-solutions}
\bibinfo{type}{Tech. Rep.},
  \bibinfo{institution}{http://www.advancedcyclotron.com/cyclotron-solutions/tr30}
  (\bibinfo{year}{2012}).

\bibitem[{\citenamefont{Adelmann et~al.}(2008)\citenamefont{Adelmann, Kraus,
  Ineichen, Russell, Bi, and Yang}}]{Adelmann:1}
\bibinfo{author}{\bibfnamefont{A.}~\bibnamefont{Adelmann}},
  \bibinfo{author}{\bibfnamefont{C.}~\bibnamefont{Kraus}},
  \bibinfo{author}{\bibfnamefont{Y.}~\bibnamefont{Ineichen}},
  \bibinfo{author}{\bibfnamefont{S.}~\bibnamefont{Russell}},
  \bibinfo{author}{\bibfnamefont{Y.~J.} \bibnamefont{Bi}}, \bibnamefont{and}
  \bibinfo{author}{\bibfnamefont{J.~J.} \bibnamefont{Yang}},
  \bibinfo{type}{Tech. Rep.} \bibinfo{number}{PSI-PR-08-02},
  \bibinfo{institution}{Paul Scherrer Institut} (\bibinfo{year}{2008}).

\bibitem[{\citenamefont{Yang et~al.}(2010)\citenamefont{Yang, Adelmann, Humbel,
  Seidel, and Zhang}}]{Yang:1}
\bibinfo{author}{\bibfnamefont{J.~J.} \bibnamefont{Yang}},
  \bibinfo{author}{\bibfnamefont{A.}~\bibnamefont{Adelmann}},
  \bibinfo{author}{\bibfnamefont{M.}~\bibnamefont{Humbel}},
  \bibinfo{author}{\bibfnamefont{M.}~\bibnamefont{Seidel}}, \bibnamefont{and}
  \bibinfo{author}{\bibfnamefont{T.~J.} \bibnamefont{Zhang}},
  \bibinfo{journal}{Phys. Rev. ST Accel. Beams} \textbf{\bibinfo{volume}{13}},
  \bibinfo{pages}{064201} (\bibinfo{year}{2010}),
  \urlprefix\url{http://link.aps.org/doi/10.1103/PhysRevSTAB.13.064201}.

\bibitem[{\citenamefont{Yang et~al.}(2011)\citenamefont{Yang, Zhang, Lin,
  Adelmann, Wang, and An}}]{Yang:2}
\bibinfo{author}{\bibfnamefont{J.~J.} \bibnamefont{Yang}},
  \bibinfo{author}{\bibfnamefont{T.~J.} \bibnamefont{Zhang}},
  \bibinfo{author}{\bibfnamefont{Y.~Z.} \bibnamefont{Lin}},
  \bibinfo{author}{\bibfnamefont{A.}~\bibnamefont{Adelmann}},
  \bibinfo{author}{\bibfnamefont{F.}~\bibnamefont{Wang}}, \bibnamefont{and}
  \bibinfo{author}{\bibfnamefont{S.~Z.} \bibnamefont{An}},
  \bibinfo{journal}{Sci China Phys Mech Astron} \textbf{\bibinfo{volume}{54
  S2}}, \bibinfo{pages}{1} (\bibinfo{year}{2011}).

\bibitem[{\citenamefont{Gordon}(1969)}]{Gordon:1}
\bibinfo{author}{\bibfnamefont{M.~M.} \bibnamefont{Gordon}}, in
  \emph{\bibinfo{booktitle}{Proc. 5th Int. Conf. on Cyclotrons and their
  Applications}} (\bibinfo{address}{Oxford}, \bibinfo{year}{1969}), p.
  \bibinfo{pages}{305}.

\bibitem[{\citenamefont{Kleeven}(1988)}]{kleeven:1}
\bibinfo{author}{\bibfnamefont{W.}~\bibnamefont{Kleeven}}, Ph.D. thesis,
  \bibinfo{school}{TU Eindhoven} (\bibinfo{year}{1988}).

\bibitem[{\citenamefont{Adam}(1985)}]{Adam:0}
\bibinfo{author}{\bibfnamefont{S.}~\bibnamefont{Adam}}, Ph.D. thesis,
  \bibinfo{school}{ETHZ, Switzerland} (\bibinfo{year}{1985}), \bibinfo{note}{no
  7694}.

\bibitem[{\citenamefont{Bertrand and Ricaud}(2001)}]{Bert:2001}
\bibinfo{author}{\bibfnamefont{P.}~\bibnamefont{Bertrand}} \bibnamefont{and}
  \bibinfo{author}{\bibfnamefont{C.}~\bibnamefont{Ricaud}}, in
  \emph{\bibinfo{booktitle}{Conf. on Cyclotrons and their Applications}}
  (\bibinfo{address}{East Lansing, Michigan}, \bibinfo{year}{2001}), p.
  \bibinfo{pages}{379}.

\bibitem[{\citenamefont{Adelmann}(2002)}]{Ada:1}
\bibinfo{author}{\bibfnamefont{A.}~\bibnamefont{Adelmann}}, Ph.D. thesis,
  \bibinfo{school}{ETHZ, Switzerland} (\bibinfo{year}{2002}), \bibinfo{note}{no
  14545}.

\bibitem[{\citenamefont{Pozdeyev et~al.}(2009)\citenamefont{Pozdeyev,
  Rodriguez, Marti, and York}}]{Pozdeyev:2}
\bibinfo{author}{\bibfnamefont{E.}~\bibnamefont{Pozdeyev}},
  \bibinfo{author}{\bibfnamefont{J.~A.} \bibnamefont{Rodriguez}},
  \bibinfo{author}{\bibfnamefont{F.}~\bibnamefont{Marti}}, \bibnamefont{and}
  \bibinfo{author}{\bibfnamefont{R.~C.} \bibnamefont{York}},
  \bibinfo{journal}{Phys. Rev. ST Accel. Beams} \textbf{\bibinfo{volume}{12}},
  \bibinfo{pages}{054202} (\bibinfo{year}{2009}),
  \urlprefix\url{http://link.aps.org/doi/10.1103/PhysRevSTAB.12.054202}.

\bibitem[{\citenamefont{Koscielniak and Adam}(1993)}]{Adam:2}
\bibinfo{author}{\bibfnamefont{S.}~\bibnamefont{Koscielniak}} \bibnamefont{and}
  \bibinfo{author}{\bibfnamefont{S.}~\bibnamefont{Adam}}, in
  \emph{\bibinfo{booktitle}{Proc. Particle Accelerator Conf.}}
  (\bibinfo{address}{Washington}, \bibinfo{year}{1993}), p.
  \bibinfo{pages}{3639}.

\bibitem[{\citenamefont{Humbel}(2009)}]{HumbPC}
\bibinfo{author}{\bibfnamefont{M.}~\bibnamefont{Humbel}},
  \bibinfo{howpublished}{private communication} (\bibinfo{year}{2009}).

\bibitem[{\citenamefont{Baumgarten}(2011)}]{Baumgarten:1}
\bibinfo{author}{\bibfnamefont{C.}~\bibnamefont{Baumgarten}},
  \bibinfo{journal}{Phys. Rev. ST Accel. Beams} \textbf{\bibinfo{volume}{14}},
  \bibinfo{pages}{114201} (\bibinfo{year}{2011}),
  \urlprefix\url{http://link.aps.org/doi/10.1103/PhysRevSTAB.14.114201}.

\bibitem[{\citenamefont{{R. Miracoli, {\it et al.}}}(2012)}]{celona-1}
\bibinfo{author}{\bibnamefont{{R. Miracoli, {\it et al.}}}},
  \bibinfo{journal}{Rev. Sci. Instrum.} \textbf{\bibinfo{volume}{83}}
  (\bibinfo{year}{2012}).

\bibitem[{\citenamefont{{F. Maimone, {\it et al.}}}(2008)}]{maimone-1}
\bibinfo{author}{\bibnamefont{{F. Maimone, {\it et al.}}}}, in
  \emph{\bibinfo{booktitle}{Proc. Eur. Part. Acc. Conf. Genoa}}
  (\bibinfo{year}{2008}), \bibinfo{note}{{MOPC151}}.

\bibitem[{\citenamefont{{A. Bungau, {\it et al.}}}(2012)}]{isodar}
\bibinfo{author}{\bibnamefont{{A. Bungau, {\it et al.}}}},
  \bibinfo{journal}{arXiv:1205.4419v1, {accepted for publication in PRL}}
  (\bibinfo{year}{2012}).

\bibitem[{\citenamefont{Bi et~al.}(2010)\citenamefont{Bi, Zhang, Tang, Huang,
  and Yang}}]{Bi:1}
\bibinfo{author}{\bibfnamefont{Y.~J.} \bibnamefont{Bi}},
  \bibinfo{author}{\bibfnamefont{T.~J.} \bibnamefont{Zhang}},
  \bibinfo{author}{\bibfnamefont{C.~X.} \bibnamefont{Tang}},
  \bibinfo{author}{\bibfnamefont{Y.~S.} \bibnamefont{Huang}}, \bibnamefont{and}
  \bibinfo{author}{\bibfnamefont{J.~J.} \bibnamefont{Yang}},
  \bibinfo{journal}{J. Appl. Phys.} \textbf{\bibinfo{volume}{107}},
  \bibinfo{pages}{063304} (\bibinfo{year}{2010}).

\end{thebibliography}
\end{document}